\documentclass[acmsmall]{acmart}

\usepackage{balance,bm}
\usepackage{xcolor,colortbl}
\usepackage{url}
\usepackage{enumitem}

\usepackage{graphicx}  
\usepackage{textcomp, soul}
\usepackage{wrapfig}
\usepackage{subcaption}
\usepackage{mathtools}
\usepackage{algorithmic}
\usepackage[ruled]{algorithm2e}
\usepackage[T1]{fontenc}
\usepackage[utf8]{inputenc}
\usepackage{multirow, verbatim, xspace}
\usepackage[many]{tcolorbox}
\usepackage{titlesec}
\usepackage{arydshln}
\usepackage{longtable}
\usepackage{booktabs}

\definecolor{linkColor}{RGB}{6,125,233}
\definecolor{green}{rgb}{0.0, 0.65, 0.31}
\definecolor{bleudefrance}{rgb}{0.19, 0.55, 0.91}
\definecolor{ceruleanblue}{rgb}{0.16, 0.32, 0.75}
\definecolor{grey}{HTML}{969696}
\definecolor{lightgrey}{HTML}{d7d7d7}
\definecolor{greybackground}{HTML}{e9ecef}
\definecolor{violet}{HTML}{6a51a3}
\definecolor{lgreen}{HTML}{5ab4ac}
\definecolor{dgreen}{HTML}{005a32}
\definecolor{purple}{HTML}{ae017e}
\definecolor{orange}{HTML}{d95f0e}

\definecolor{generate}{HTML}{9e3dc5}
\definecolor{understand}{HTML}{3dc40f}
\definecolor{evaluate}{HTML}{ff733c}

\definecolor{individual1}{HTML}{F5E4E4}
\definecolor{individual2}{HTML}{C95E5E}
\definecolor{care1}{HTML}{CEE9D9}
\definecolor{care2}{HTML}{417C59}
\definecolor{info1}{HTML}{EFDFF3}
\definecolor{info2}{HTML}{9249A3}
\definecolor{tech1}{HTML}{D9E2FC}
\definecolor{tech2}{HTML}{435993}

\colorlet{tableheadcolor}{gray!25} 
\colorlet{tablerowcolor}{gray!15} 
\colorlet{tablerowcolor2}{gray!12} 
\colorlet{tablerowcolor3}{gray!25} 

\newcommand{\rowcollight}{\rowcolor{tablerowcolor2}} %

\renewcommand{\textrightarrow}{$\rightarrow$}

\colorlet{tableheadcolor}{gray!25} 
\colorlet{tablerowcolor}{gray!5} 

\newtcolorbox{roundBox}{
  enhanced,
  breakable,
  before skip = 0.5em,
  after skip = 0.5em, 
  top = 0.3em,
  bottom = 0.3em,
  colback = grey!10, 
  boxrule = 0pt,
  rounded corners,
  arc = 3pt,
  fontupper= \small\itshape
}

\newtcolorbox{featureBox}{
  enhanced,
  breakable,
  before skip = 0.5em,
  after skip = 0.5em, 
  top = 0.3em,
  bottom = 0.3em,
  colback = greybackground, 
  boxrule = 0pt,
  rounded corners,
  arc = 3pt,
}

\AtBeginDocument{%
  \providecommand\BibTeX{{%
    \normalfont B\kern-0.5em{\scshape i\kern-0.25em b}\kern-0.8em\TeX}}}

\setcopyright{cc}
\setcctype{by}
\acmJournal{PACMHCI}
\acmYear{2025} \acmVolume{9} \acmNumber{7} \acmArticle{CSCW363} \acmMonth{11} \acmPrice{}\acmDOI{10.1145/3757544}

\begin{document}

\title{A Risk Taxonomy and Reflection Tool for Large Language Model Adoption in Public Health}

\author{Jiawei Zhou}
\affiliation{
  \institution{Georgia Institute of Technology}
  \city{Atlanta}
  \state{GA}
  \country{USA}
}
\email{j.zhou@gatech.edu}

\author{Amy Z. Chen}
\affiliation{
  \institution{Georgia Institute of Technology}
  \city{Atlanta}
  \state{GA}
  \country{USA}
}
\email{amychen@gatech.edu}

\author{Darshi Shah}
\affiliation{
  \institution{Georgia Institute of Technology}
  \city{Atlanta}
  \state{GA}
  \country{USA}
}
\email{dshah435@gatech.edu}

\author{Laura M. Schwab-Reese}
\affiliation{
  \institution{Purdue University}
  \city{West Lafayette}
  \state{IN}
  \country{USA}
}
\email{lschwabr@purdue.edu}

\author{Munmun De Choudhury}
\affiliation{
  \institution{Georgia Institute of Technology}
  \city{Atlanta}
  \state{GA}
  \country{USA}
}
\email{munmund@gatech.edu}

\renewcommand{\shortauthors}{Jiawei Zhou, Amy Z. Chen, Darshi Shah, Laura M. Schwab-Reese, \& Munmun De Choudhury}

\begin{abstract}  

Recent breakthroughs in large language models (LLMs) have generated both interest and concern about their potential adoption as information sources or communication tools across different domains. In public health, where stakes are high and impacts extend across diverse populations, adopting LLMs poses unique challenges that require thorough evaluation. However, structured approaches for assessing potential risks in public health remain under-explored. To address this gap, we conducted focus groups with public health professionals and individuals with lived experience to unpack their concerns, situated across three distinct and critical public health issues that demand high-quality information: infectious disease prevention (vaccines), chronic and well-being care (opioid use disorder), and community health and safety (intimate partner violence). We synthesize participants' perspectives into a risk taxonomy, identifying and contextualizing the potential harms LLMs may introduce when positioned alongside traditional health communication. This taxonomy highlights four dimensions of risk to individuals, human-centered care, information ecosystems, and technology accountability. For each dimension, we unpack specific risks and offer example reflection questions to help practitioners adopt a risk-reflexive approach. By summarizing distinctive LLM characteristics and linking them to identified risks, we discuss the need to revisit prior mental models of information behaviors and complement evaluations with external validity and domain expertise through lived experience and real-world practices. Together, this work contributes a shared vocabulary and reflection tool for people in both computing and public health to collaboratively reflect and assess risks in deciding when to employ LLM capabilities (or not) and how to mitigate harm.

\end{abstract}



\maketitle

\section{Introduction} 

\begin{quote}
    \textit{``I love new technologies but this one has me scared. I myself wouldn't feel like I know enough to ask the right questions and to evaluate the impact of it.''} --- E5 (public health director)
\end{quote}

Recent breakthroughs in large language models (LLMs) have spurred widespread attention and rapid adoption across different fields. With their abilities to generate persuasive human-like language on the basis of large sets of human-written content~\cite{Brown2020LanguageLearners,vaswani2017attention}, LLMs hold the potential to influence how we interact with information. LLMs have quickly gathered hundreds of millions of active users~\cite{Reuters2024openai} and made their way into everyday products, sometimes even without users' full awareness. Accordingly, public discourse has shown excitement about LLM assistance in various informational tasks and even hype about the framing of LLMs as the ``next-generation search engine''\cite{ico_search} or ``alternative to human therapists''~\cite{Lawrie2025AItherapists}.

This technology enthusiasm extends into public health, a public sector that aims to protect and improve the health of people and their communities by promoting healthy lifestyles, researching disease and injury prevention, and detecting, preventing and responding to infectious diseases~\cite{cdc_publichealth, turnock2009public}. Studies have presented evidence of LLMs' potential in assisting collection, analysis, and dissemination of public health information~\cite{xie2025leveraging, karinshak2023working, joshi2024ensuring}, developing public health policies~\cite{espinosa2024use, hou2025can, guevara2024large}, scaling care to population~\cite{ayers2023comparing, jo2023understanding}, and supporting community health workers~\cite{ramjee2024ashabot}. In practice, LLMs have already been implemented in community- or population-level systems or technologies, such as health record infrastructure~\cite{Landi2024epic}, virtual agents for dealing with domestic violence~\cite{mccall2023ai}, and public health monitoring instruments~\cite{jo2023understanding}. At the same time, public-facing LLM products have become widely accessible to individuals to seek support, sometimes knowingly and other times embedded within existing platforms~\cite{AlSibai2025ChatGPTMeds, McMahon2024Glue}.

While this proliferation seems promising, experts and scholars caution against exaggerating LLM capabilities~\cite{drogt2024ethical, wornow2023shaky}. Concerns have been raised about the potential of generating harmful content~\cite{zhou2023synthetic, liang2022holistic, gehman2020realtoxicityprompts}, perpetuating biases~\cite{omiye2023large}, compromising data privacy~\cite{zhang2024s}, limited generalizability of models~\cite{wornow2023shaky}, and challenges in real-world implementation~\cite{wester2024chatbot, jo2023understanding}. Given the high-stakes nature and data sensitivity of population-level health, it deserves careful consideration of when it is appropriate to employ LLM capacities and how we can mitigate potential harm when we do. 

Yet, despite this recognized need for caution, evaluating the risks of LLM adoption in public health real practice remains a complex challenge. The first gap lies in the lack of domain understanding, as the majority of risk taxonomies lack granularity to be applied for specific uses~\cite{liao2023ai} or are created within the computer science community and tend to leave out domain experts and real users~\cite{shelby2023sociotechnical, weidinger2022taxonomy, bender2021dangers}. Second, when considering population-level impact, potential risks come in ecological ``layers'' from individuals to society, and conventional categorizations based on content types (e.g., misinformation, hate speech, and biases) are not sufficient to evaluate real-world cases~\cite{scheuerman2021framework, zhou2025harm}. Identifying the presence of generated harmful content is only the first step; we also need to contextualize the documented possibility of low-quality information within specific public health issues and relevant populations to evaluate the consequences and mitigate potential harm. In response to these gaps, this paper examines risks associated with the use of LLMs to offer and seek health informational assistance within the public health domain. We ground our analysis in three distinct and critical public health issues, examining them through the perspectives of both public health professionals and individuals with lived experience who actively seek online health information. Specifically, we ask: \textit{What negative consequences might arise from adopting LLMs to meet informational needs in public health?}

We conducted focus group sessions with ten public health professionals and ten individuals with lived experience who have actively sought online health information, in order to jointly explore potential negative influences of using LLMs in public health. We selected vaccines, opioid use disorder, and intimate partner violence as topics for different sessions based on their significance across different dimensions of public health: infectious disease prevention, chronic and well-being care, and community health and safety --- all demanding high-quality information with existing prevalent issues such as misinformation, biases, and sensitivity. The result is a risk taxonomy of potential adoption of LLM for public health. Our taxonomy consists of four dimensions of harm: risks to individuals, human-centered care, information ecosystems, and technology accountability (Fig~\ref{fig:overview}). Within each area, we list specific risks and associated example reflection questions to help practitioners in both computing and health fields become reflexive and risk-aware.

This work makes two main contributions. \textbf{(1)} It gives a comprehensive and grounded list of possible risks in implementing LLMs for public health, and differs from prior generic taxonomies by being rooted in public health issues and informed by both professional and lived experience. To our knowledge, this is the first work to comprehensively explore risks of LLMs for public health. \textbf{(2)} It offers a shared vocabulary through lists of LLM distinctive characteristics (Fig~\ref{fig:feature}) and types of LLM-generated low-quality information (Appendix~\ref{appendix:low_quality}), a risk taxonomy (Fig~\ref{fig:overview}), as well as reflection questions for operationalization (Sec~\ref{sec:reflection_question}). These tools can help future collaborations between experts in computing and health fields to promote careful considerations of adoption consequences.

\vspace{0.5em}
\noindent \textbf{Content Warning:} We caution readers that this paper discusses sensitive topics, including intimate partner violence and opioid use disorder. Some readers may find certain quotes and descriptions to be emotionally triggering.

\section{Related Work} 

\subsection{Digital Support and Threats to Health Information}
The integration of information and communication technology has transformed how care seekers and providers seek, access, and interact with health information. About 59\% of adults in the U.S. have searched for health or medical information online in the past 12 months~\cite{wang2023health}. Notable technology examples include search engines, social media platforms, and online communities, as well as applications for health education and intervention~\cite{neuhauser2003rethinking, de2014seeking, chen2025caring}. Various types of digital tools democratize health information access and authorship~\cite{zhao2017consumer, costello2016impact}, provide safe spaces for individuals facing stigma or legal concerns~\cite{zhao2017consumer, johnson2006neo, de2014seeking}, and connect people with individuals who share similar experiences~\cite{augustaitis2021online, jacobs2019think, zhou2022veteran}.

Despite the accessibility and benefits, finding high-quality health information and protecting privacy remains a challenge. For example, \citet{hansen2003adolescents} found that only 69\% of observed searches in their study resulted in a correct and useful answer. Online users, especially younger generations, do not always verify sources or assess information reliability~\cite{hansen2003adolescents, hassoun2023practicing}, while echo chambers can further amplify biased perspectives~\cite{jiang2021social}. Online platforms can facilitate the spread of inaccurate health information, whether shared innocently or with deceptive intent, and the prevalence of low-quality information can gradually erode trust and discourage online health-seeking behaviors~\cite{zhao2017consumer, augustaitis2021online}. Privacy concerns also emerge. While online anonymity allows individuals to seek sensitive health information~\cite{andalibi2016understanding}, it also poses risks of individuals being identified when their data is aggregated across multiple online sources~\cite{li2011new}.

The nature of digital interactions presents additional concerns in further marginalizing, isolating, or even silencing individuals. First, disparities in digital access and literacy can contribute to marginalization. Individuals with lower health literacy face greater challenges in finding, evaluating, and understanding online health information~\cite{manganello2017relationship}. Internet access is also unevenly distributed, with fewer online health information seekers in developing regions compared to developed ones~\cite{jia2021online}. Meanwhile, online platforms, despite being a ``safe space'' to some, can lack the interpersonal trust-building processes inherent in traditional healthcare interactions~\cite{johnson2006neo}. In some cases, technology can even be misused to silence marginalized voices rather than empower them~\cite{foriest2024cross}. 

These challenges become even more pronounced with the rise of generative AI that can rapidly produce human-like content at scale with inconsistent quality~\cite{liang2022holistic, chang2024survey, zhou2023synthetic} and potential influences on people's critical thinking~\cite{lee2025impact}. While prior work has examined various digital threats to health information, there remains a gap in capturing changes brought by AI's generative abilities and comprehensively understanding the potential negative consequences of LLMs in this area. Our study builds on prior scholarship and addresses this gap by situating LLMs within representative public health issues, positioning them alongside traditional health communication, and engaging individuals outside the computing field who bring professional knowledge or lived experience.

\subsection{Potential and Challenges of LLMs in Public Health}
A growing body of work has explored the potential of large language models in supporting the three core goals of public health agencies: assessment, policy development, and assurance~\cite{turnock2009public, division1988future}. In assisting systematic collection, analysis, and dissemination of health information, LLMs can help assisting infectious disease surveillance~\cite{xie2025leveraging}, promoting health interventions and healthy lifestyles~\cite{karinshak2023working}, and combating misinformation or myths in facing public health crisis~\cite{joshi2024ensuring, mittal2025onine}. In supporting the development of public health policies, LLMs can be used to understand public health discourse~\cite{espinosa2024use}, while generative agents can also simulate human behavior to inform public health policy~\cite{hou2025can}. In assuring delivering personal and community-wide health services to every member of the community, LLMs hold the potential to scale provider to populations~\cite{ayers2023comparing, jo2023understanding}, monitor social determinants of health (SDoH)~\cite{guevara2024large}, as well as support community health workers~\cite{ramjee2024ashabot}. In practice, LLMs have already been implemented in community- or population-level systems or technologies, such as health record infrastructure~\cite{Landi2024epic}, virtual agents for dealing with domestic violence~\cite{mccall2023ai}, and public health monitoring instruments~\cite{jo2023understanding}.

Orthogonally, research has also highlighted challenges for adopting LLMs for informational needs and public health support. When providing informational assistance, LLMs face documented issues in data privacy~\cite{zhang2024s}, language and geographic disparities~\cite{jin2024ask, moayeri2024worldbench}, and creation of inaccurate, biased, or toxic content~\cite{zhou2023synthetic, magu2025navigating, chandra2024lived, yoo2025ai}. These limitations could lead to exaggerated or new risks in public health, a field that often involves both sensitive or high-stakes issues and serves diverse populations~\cite{turnock2009public}. In health contexts, there remain questions about LLMs' ability to provide reliable and effective information. For instance, \citet{harrer2023attention} found that only half of the LLM-generated messages met the standards for direct adoption for use. When used for communication assistance, LLMs are found to have difficulty in being customized or controlled~\cite{wester2024chatbot} and a lack of sensitivity in managing emotional needs and emergencies~\cite{wang2021evaluation}. 
Tailoring language to suit diverse demographics is another challenge~\cite{jo2023understanding}, particularly for marginalized groups, such as the LGBTQ+ community~\cite{ma2024evaluating, ovalle2023m} and people with disabilities~\cite{gadiraju2023wouldn}, where cultural nuances and sensitive topics may be overlooked. In addition, a core goal of public health is to reduce health disparities and deliver interventions at the population level~\cite{turnock2009public}. However, the tendency of LLMs to propagate inaccurate and biased race-based suggestions~\cite{omiye2023large} may actually hinder broader public health efforts. 

Accordingly, scholarship has started to critically reflect on the way we approach LLM evaluation -- an effort of significant importance to public health that is grounded in real practice and community efforts. Researchers have called for better reporting transparency with contextualized and empirically grounded evidence and claims~\cite{drogt2024ethical}, assessment reliability that accounts for practical deployment and data representation~\cite{wornow2023shaky}, diverse assessment methodologies and raters in surfacing health equity-related biases~\cite{pfohl2024toolbox}, and attention to patient-provider dynamics in practice~\cite{wilcox2023ai}. For instance, \citet{wornow2023shaky} found in their survey study that the generalizability of most clinical foundational models was limited by an overreliance on structured codes and non-representative datasets. At the same time, most evaluations do not offer meaningful insights for real-world applications, instead focusing on traditional natural language processing (NLP) tasks~\cite{wornow2023shaky}. As a result, they advocate for frameworks that align more closely with values and factors crucial to health contexts. We build on prior scholarship by comprehensively examining risks posed by LLMs in public health through three critical yet distinctive issues.

\subsection{AI and LLMs Risk Taxonomies}

In response to the growing literature on the ethics and limitations of AI and LLMs, researchers have proposed various taxonomies to address risks posed by these models. Some specifically examine the potential risks of LLMs but take a general perspective on their impacts. One of the earliest efforts is the position paper by \citet{bender2021dangers} that highlights the dangers in environmental and financial costs, overrepresentation of hegemonic viewpoints and encoded biases, hype or misunderstanding in language model capabilities, and misconception of synthetic text as meaningful. In a more structured approach, \citet{weidinger2022taxonomy} summarized six risk areas based on computing expert discussion and supporting literature, which include discrimination, hate speech and exclusion, information hazards, misinformation harms, malicious uses, human-computer interaction harms, and environmental and socioeconomic harms. Other studies adopt a more focused view on specific areas of impact. For instance, \citet{shelby2023sociotechnical} conducted a scoping review on sociotechnical harm of algorithmic systems and identified five major themes: representational, allocative, quality-of-service, interpersonal harms, and social system/societal harms. Similarly, \citet{dominguez2024mapping} conducted a review using snowball and structured search to study the social, political, and environmental impacts of foundation models, which leads to 14 categories of risks --- including less-studied ones of unreliable performance, lock-in and opacity risks, and value misalignment. In the realm of health, \citet{de2023benefits} discussed the benefits and harms of LLMs in digital mental health across four application areas: supporting individual care-seeking, assisting caregiving, decision-making aid, and transforming telehealth paradigms. \citet{hamid2024review} reviewed prior literature and summarized seven open research challenges: misinformation, resource limitations, bias and fairness in healthcare LLMs, interpretability and transparency, integration within clinical workflows, ethical considerations and patient privacy, and clinical validation and real-world performance.

Despite these contributions, there is no established vocabulary to specifically map the risks in public health contexts. Additionally, public health encompasses a broad scope for care-seeking and -providing that extends beyond health encounters, involving a wide range of stakeholders such as social workers, researchers, and community planners. However, the expertise built through professional practice and lived experience remains largely missing from current discussions, despite public health's emphasis on real-world practice and field involvement~\cite{turnock2009public, cdc_publichealth}. This absence of a shared framework limits our ability to systematically explore, communicate, and mitigate potential harms in this specific domain.

From a methodological point, most existing taxonomies are position papers or review studies centered within the computer science field, which often exclude domain experts and end users in applied areas. Research has warned of the limitations of relying solely on NLP metrics and expertise~\cite{wornow2023shaky, pfohl2024toolbox}. Consequently, research in human-computer interaction (HCI) and computer-supported cooperative work (CSCW) has called for human-centered approaches and stakeholder involvement, as certain harms may be difficult to capture solely through existing AI literature~\cite{wilcox2023ai, antoniak2024nlp, kawakami2024situate}. Some recent efforts in health include the work by \citet{antoniak2024nlp} that involved diverse stakeholders, including healthcare workers and birthing people, to propose practical principles for ethical use of NLP and LLMs for maternal health. Our work seeks to address the gap in public health contexts by proposing a shared vocabulary grounded in public health issues, informed by perspectives of both public health professionals and individuals with lived experience.

\section{Method} 

\subsection{Public Health Topic Selection and Context}
To contextualize and evaluate the risks of using large language models (LLMs) for informational needs in public health, we chose three distinct and critical issues: vaccines, opioid use disorder (OUD), and intimate partner violence (IPV). The selection of these topics was driven by their significance across different dimensions of public health --- infectious disease prevention, chronic and well-being care, and community health and safety. Each topic underscores the importance of high-quality information in public health communication, which often involves both sensitive or high-stakes issues and diverse populations. 
Specifically, vaccines play a crucial role in infectious disease prevention, while misinformation and public distrust significantly contribute to vaccine hesitancy~\cite{pierri2022online, puri2020social, diamond2022polyvocality}. Effective counter-speech is needed to combat these misconceptions and enhance trust~\cite{zhang2021cultural, zimet2013beliefs}. OUD is a major issue in chronic and well-being care, which is exacerbated by stigma~\cite{elsherief2024identification, olsen2014confronting} and vulnerability~\cite{yamamoto2019association, van2020socioeconomic}. This health crisis needs public education to reduce biases and create a supportive environment. IPV is a high-risk and stigmatized issue in community health and safety that demands highly sensitive and supportive communication~\cite{schwab2017attitudinal, mittal2024news}. IPV involves complex challenges, including personal safety, mental health, and legal and financial challenges~\cite{schwab2016associations, freed2017digital}. Providing high-quality, empathetic information is essential for supporting survivors and addressing their comprehensive needs.

\subsection{Participants and Recruitment}

We recruited both public health professionals and members of the general public to reflect on how LLMs could be used to provide and seek help in public health. Health professionals utilize information technologies to support the informational needs of individuals, while individuals' independent engagement with technologies can give rise to broader behavioral patterns that professionals must anticipate and respond to. Including both groups allowed us to capture complementary perspectives on the risks and implications of LLM adoption. The health professional expert group (labeled with ``expert/E'') consisted of people working in the public health sector, such as social workers, nurses, community health workers, and researchers, all of whom have experience working on at least one of the three topics. The general public group (labeled with ``public/P'') includes people who have actively sought information about one of these three issues. Additionally, participants in the sessions of OUD and IPV are individuals who have lived experience with these issues. 

To recruit health professionals, we employed a snowball sampling approach by reaching out to professional networks and relevant organizations via email. Our primary sources were public health departments and professional organizations. We first identified programs and divisions within federal and state public health departments (e.g., immunization divisions, overdose prevention initiatives, and violence prevention initiatives), and contacted divisions across various states, including Georgia, Indiana, California, Minnesota, and Texas, to get a broad geographic coverage. During this process, we identified partner alliances and support systems listed by state departments and subsequently reached out to these organizations and shelters. Simultaneously, we distributed our recruitment flyers via professional networks and working groups, contacted ethics teams and task forces in hospitals and universities, and reached out to researchers who work on health communication and support on the chosen health issues. 

To recruit the general public, we used the Prolific~\footnote{https://www.prolific.com/} research platform, where we launched three separate studies for each of our health topics. For OUD and IPV sessions, we set Prolific's screeners to include only individuals who have experienced substance abuse problems or abuse-related incidents, and specified eligibility for those who have personally experienced OUD or IPV. We did not collect any personally identifiable information, and all communications were carried on the Prolific platform. All public participants' responses to the 5-minute screening survey were rewarded with an hourly rate of \$12. 

For both participant groups, inclusion criteria were adults aged 18 or older, residing in the U.S., and having access to a digital device to join an online meeting and participate in brainstorming and writing activities. Interested respondents completed a screening survey about their demographics, occupation, professional background (for the expert cohort), AI attitudes, familiarity with LLM products, and their availability to suggested focus group timings. Participants were selected for focus groups solely based on their availability.

\begin{table}[t]
\centering
\sffamily
\small
\caption{Participant information. Experts (E1-10) are public health professionals who engage with community-based health intervention or prevention efforts and have created or shared information about selected topics. Experts' years of professional experience in occupations are indicated in brackets. Public individuals (P1-10) are people who have experienced questions about selected topics and actively sought online information. Public participants in OUD and IPV sessions have lived experiences with these issues.}
\label{table:participant}
\resizebox{\columnwidth}{!}{%
\begin{tabular}{lllllllll} 
    \textbf{ID} & \textbf{Occupation(yrs)} & \textbf{Topic} & \textbf{Age} & \textbf{Gender} & \textbf{Ethnicity} & \textbf{Education} & \textbf{Experience with AI tools} & \textbf{Use of LLM tools}\\
    \toprule

    
    E1 & Social worker(18) & IPV & 35-44 & Female & White & Advanced degree & Only read about it & Not heard about it\\ 

    E2 & Researcher(14) & IPV & 45-54 & Female & White & Advanced degree & Worked on related topics & Sometimes, but not regularly\\ 

    E3 & Social worker(13) & IPV & 35-44 & Female & White & Advanced degree & Only read about it & Never used, but heard about it\\ 

    E4 & Researcher(4) & IPV & 18-24 & Female & Asian & Bachelor's degree & Occasionally used AI tools & Sometimes, but not regularly\\ 

    
    \rowcollight P1 & Employed & IPV & 35-44 & Female & White & Advanced degree & Occasionally used AI tools & Rarely, use only occasionally\\ 

    \rowcollight P2 & On Leave & IPV & 55-64 & Female & White & Associate degree & Extensively used AI tools & Always, use whenever applicable\\ 

    \rowcollight P3 & Unemployed & IPV & 35-44 & Female & Black/African Ame & Some college, no degree & Occasionally used AI tools & Sometimes, but not regularly\\ 

    \hline


    E5 & Public health director(30) & OUD & 65+ & Female & White & Advanced degree & Worked on related topics & Sometimes, but not regularly\\ 

    E6 & Nurse(5) & OUD & 35-44 & Non-binary & White & Advanced degree & Occasionally used AI tools & Often, as part of my routine\\ 

    E7 & Nurse(5) & OUD & 45-54 & Female & Hispanic/Latina & Advanced degree & In-depth understanding & Often, as part of my routine\\ 

    
    \rowcollight P4 & Self-employed & OUD & 35-44 & Female & White & Bachelor's degree & Occasionally used AI tools & Sometimes, but not regularly\\ 

    \rowcollight P5 & Student & OUD & 25-34 & Female & White & Some college, no degree & In-depth understanding & Sometimes, but not regularly\\ 

    \rowcollight P6 & Employed & OUD & 35-44 & Female & White & Bachelor's degree & Extensively used AI tools & Always, use whenever applicable\\ 

    \rowcollight P7 & Employed & OUD & 35-44 & Male & White & Some college, no degree & Extensively used AI tools & Always, use whenever applicable\\ 
    
    \hline

    E8 & Public health staff(35) & Vacc & 55-64 & Female & White & Advanced degree & Only read about it & Never used, but heard about it\\ 

    E9 & Coalition staff(9) & Vacc & 18-24 & Male & White & Bachelor's degree & In-depth understanding & Always, use whenever applicable\\ 

    E10 & Nurse(18) & Vacc & 45-54 & Female & Other & Advanced degree & Extensively used AI tools & Often, as part of my routine\\ 

    
    \rowcollight P8 & Student & Vacc & 18-24 & Female & Black/African American & Associate degree & Occasionally used AI tools & Always, use whenever applicable\\ 

    \rowcollight P9 & Employed & Vacc & 25-34 & Female & Black/African American & Some college, no degree & Occasionally used AI tools & Rarely, use only occasionally\\ 

    \rowcollight P10 & Employed & Vacc & 25-34 & Female & Black/African American & Advanced degree & Occasionally used AI tools & Rarely, use only occasionally \\ 
    
    
    \bottomrule

    \end{tabular}
}
\end{table}

In total, ten public health professionals (E1-10) and ten individuals who have experienced questions about selected topics (P1-10) participated in focus group sessions. E5 participated in an interview format after two other participants dropped out at the last minute due to medical emergencies. We compensated \$75 and \$30 for experts and public participants respectively for their time. On average, our participants have a moderate level of previous experience in AI tools and usage of LLM tools (2.7 and 3.15 on a [0,5] scale), with health professionals having slightly more experience in AI but less use of LLM. Health professionals have a median of 13.5 years of experience in their job. Table~\ref{table:participant} provides a summary of participant information.

\subsection{Study Procedure}

\begin{figure}
    \centering
    \includegraphics[width=\linewidth]{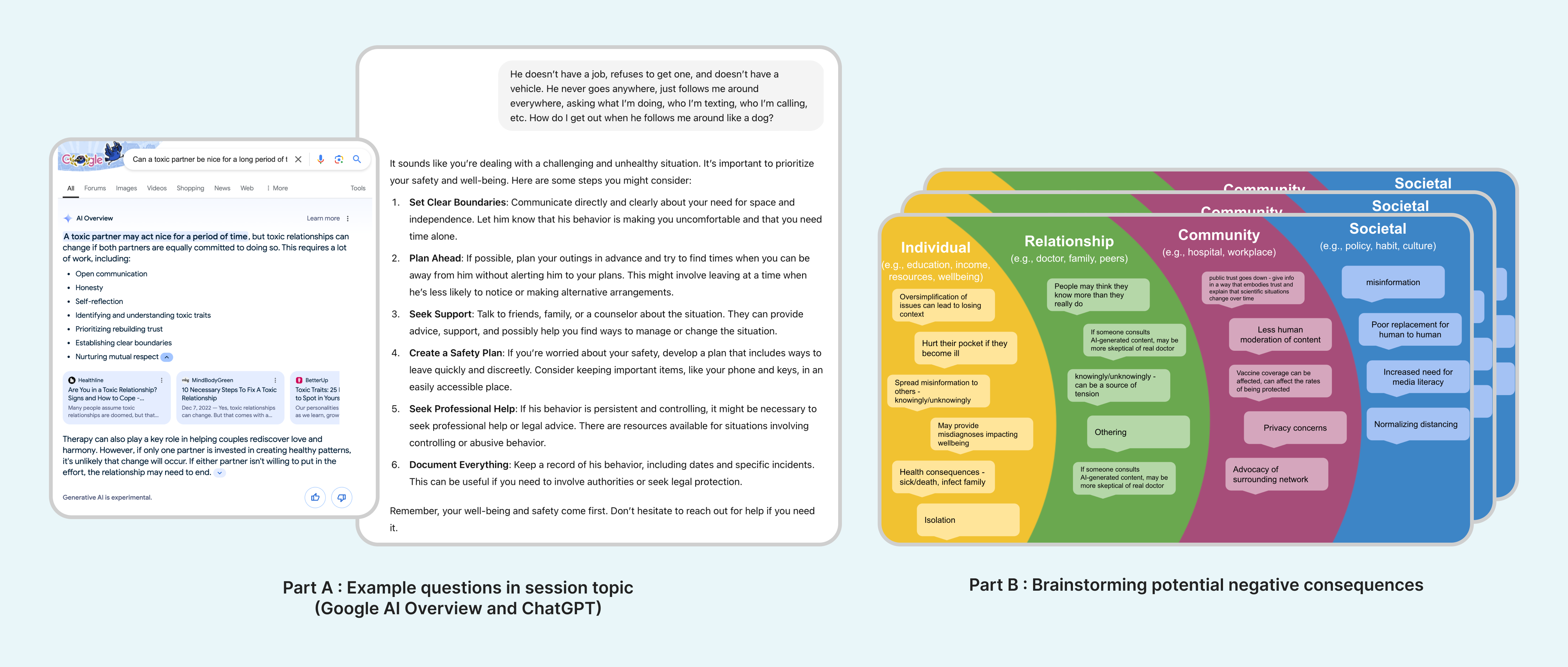}
    \caption{Some activities included in the focus group sessions: (A) Demonstrating LLMs' interaction mode and response quality through example questions on the session topic, (B) Brainstorming potential negative consequences of using LLM for public health.}
    \label{fig:activity}
\end{figure}

Due to the intentionally wide geographic distribution of our participants, all focus group sessions were conducted virtually on Zoom and lasted 90 minutes. Before the actual sessions, we conducted two rounds of pilot tests with four domain experts and four general participants to refine the study design. To provide a safe and comfortable environment for our participants, we held separate sessions for professionals and experiencers to ensure that experiencers participated in sessions with individuals who shared similar lived experiences, and encouraged participants to use any preferred name during the session. 

Each session began with an introduction to the study's goals and process, a reminder of participant rights, and self-introductions where participants shared their relevant experience with the session's topic. To contextualize the discussion of the potential role of LLMs in meeting informational needs, we started by discussing participants' experience and practices in meeting informational needs, either as information seekers or creators. Specifically, we asked about how they currently share or seek health information, how they evaluate the quality of content, the precautions required in sharing information or common challenges in information searching, and prevalent misconceptions about the specific topics. 

Following the discussion of participants' current practices in meeting informational needs, we gave a brief introduction to LLM mechanisms, underlining the fact that LLMs generate responses based on probabilities and are designed to produce human-like text but not necessarily accurate information, and encouraged participants to ask questions about LLMs. Considering the varying levels of prior experience with LLMs, we then demonstrated their interaction mode and response quality by testing two example questions relevant to the session topic (Fig.~\ref{fig:activity} Part A) while encouraging participants to try out their own questions as well. These example questions were paraphrased from online communities of r/abusiverelationships, r/OpiatesRecovery, and r/HPV~\footnote{We used HPV vaccines as a specific example of vaccines to generate LLM responses.} on Reddit. To ensure consistency across sessions, we recorded the screen while logged out and played recordings during the session. Example questions were separately tested on ChatGPT (GPT-4o-mini) and Google AI Overview (Gemini 1.5~\footnote{While at the time of this study, Google did not disclose model specifics of Google AI Overview, it was likely to be Gemini 1.5 based on the Google Search team's announcement blog (https://blog.google/products/search/ai-mode-search/).}) to show different interaction styles. We emphasized to participants that the goal was not to compare these tools directly, as questions were inherently different.

After establishing a basic understanding of LLMs and their capabilities, we introduced the final activity where participants brainstormed potential negative consequences of using LLMs for health informational needs (Fig.~\ref{fig:activity} Part B). Before the brainstorming, we introduced the socio-ecological model~\cite{bronfenbrenner1994ecological}, a framework commonly used in public health~\cite{centers2015models}, to assist participants in thinking about impacts at different levels: individual (e.g., education, resources), relationship (e.g., physician, family), community (e.g., hospital, workplace), and society (e.g., policy, habits). Participants first wrote down potential consequences individually (if the session had three participants) or in breakout groups (if the session had four participants), and then discussed with the larger group.

\subsection{Data Analysis}
We identified the types of risks raised by participants by analyzing session transcripts and brainstorming notes through a grounded approach using reflexive thematic analysis~\cite{braun2006using, braun2019reflecting, braun2021one}. Three researchers independently performed open coding on three sessions in a line-by-line fashion to identify mentioned risks, and each identified lower-level codes. These initial codes were formed verbatim from participants' conversations or brainstorming entries (e.g., \textit{``more reliance on tech vs. human relationships''}) or interpretively coded from the discussion (e.g., \textit{``don't understand prompt creation enough to use it effectively''}). Researchers then met to discuss their interpretations and consolidate the codes to reach a consensus. The first author proceeded to code the remaining three sessions and discussed any new codes with the research team. Then, researchers independently grouped the codes into higher-level themes and discussed them to reach a revised set of themes reflecting joint views. In distilling themes, we initially used a deductive approach based on the socio-ecological model's themes~\cite{bronfenbrenner1994ecological}, but ultimately adopted an inductive approach to account for the overlapping nature of many risks and re-grouped codes into our final themes.
This collective reconciliation process was iterated several times before reaching four overarching themes that constitute the dimensions of our risk taxonomy, with 11 sub-types of risks. These themes were summarized, documented, and later shared with two senior researchers for feedback and potential refinement. The final themes and findings are presented below.

\subsection{Privacy, Ethics, and Disclosure}
We understand the sensitivity of health topics in this study and potential concerns for safety and privacy, and we are committed to ensuring the privacy and safety of our participants. This study was approved by the Institutional Review Board (IRB) at our institution. The demographic information and video recordings were collected with consent and later anonymized. Any personal information, such as locations and workplaces, was removed. We refrained from collecting any personally identifiable information from people with lived experience, and data from screened-out or dropped-out participants was discarded.
Throughout the recruitment and focus groups, we assured the participants that their participation was completely voluntary, all questions were optional, and their responses would be anonymous. We also requested participants not to take screenshots or share what other participants said outside of the discussion. 
Our research team comprises researchers with diverse demographic and cultural backgrounds, bringing together interdisciplinary expertise in HCI, CSCW, and public health. Our collective experience includes in-depth work on topics such as health misinformation, mental health, and violence.

\section{Findings: Risk Taxonomy for LLM Adoption in Public Health}

\begin{figure}[h!]
    \centering
    \includegraphics[width=0.8\linewidth]{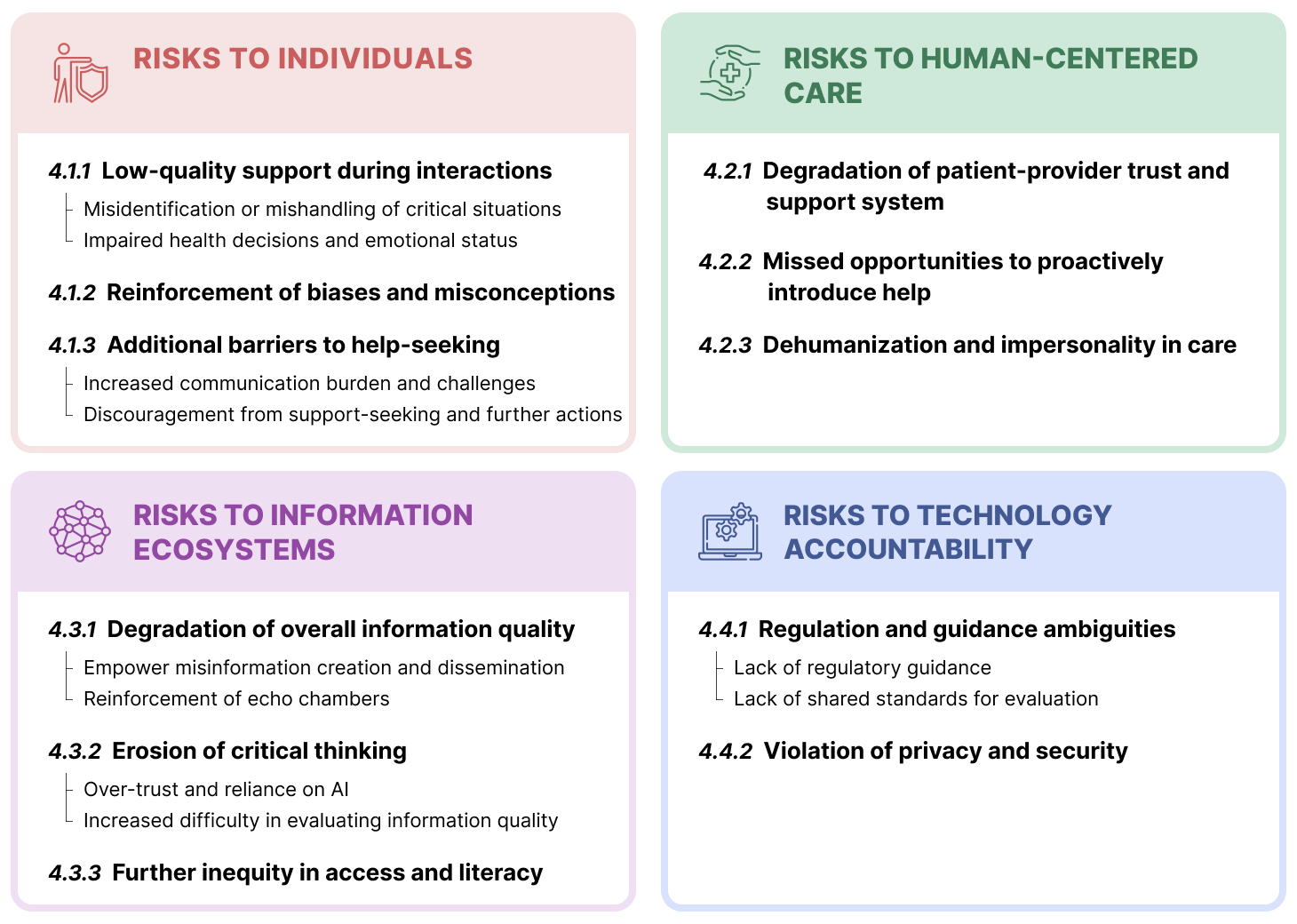}
    \caption{We summarized four dimensions of risks that LLMs pose to public health: \textcolor{individual2}{(1)} Risks to Individuals, \textcolor{care2}{(2)} Risks to Human-Centered Care, \textcolor{info2}{(3)} Risks to Information Ecosystems, \textcolor{tech2}{(4)} Risks to Technology Accountability.}
    \label{fig:overview}
\end{figure}

In anticipating negative consequences, we distill four dimensions of risks when adopting LLMs for public health through thematic analysis of session transcripts and brainstorming notes. As shown in Fig.~\ref{fig:overview}, the four dimensions include risks to (1) individuals, (2) human-centered care, (3) information ecosystems, and (4) technology accountability. From Sec~\ref{findings-individual} to Sec~\ref{findings-tech}, we unpack specific risks for each dimension. As this work has made extensive notations on low-quality information generated by LLMs, we have also included in the appendix~\ref{appendix:low_quality} a consolidated list of the different types of low-quality content identified by our participants.

While a number of risks were identified, it is important to recognize that public health serves broad and diverse populations, and the emphasis on risks will naturally vary. Across focus group sessions, participants highlighted different types of risks based on the unique needs and challenges of each topic. Specifically, in the context of vaccines, where misinformation is rampant and public trust is already fragile~\cite{pierri2022online, puri2020social, diamond2022polyvocality}, participants highlighted the risk of empowerment of misinformation creation and lowering overall information quality. With OUD, where reliable and sensitive information is crucial~\cite{elsherief2024identification, olsen2014confronting}, the lack of personalized content was seen as a greater threat as it can compromise health outcomes and even cause death while discouraging support-seeking attempts with exacerbated stigma. IPV, on the other hand, is a highly relationship-based and sensitive matter~\cite{schwab2016associations, freed2017digital}; participants focused more on risks to individual safety and well-being, as well as concerns about the erosion of social support systems.

Overall, individuals with lived experience expressed more positive attitudes towards LLMs than health professionals and expressed mixed feelings about the benefit-risk tradeoffs. They more frequently pointed out the benefits of accessible information at both resource and emotional levels --- particularly for individuals without health insurance or who are emotionally burdened by social interactions --- and the possibility of lowered healthcare costs. Public health professionals, on the other hand, tended to highlight more on the human element of care in their professions, such as building relationships and cultural sensitivity with clients, as well as implications for broader health and social services systems.

\subsection{\colorbox{individual1}{Dimension 1: Risks to Individuals}}
\label{findings-individual}

Participants reflected that individuals using LLMs to seek health information may experience harm during and after the interaction. They pointed out the potential of 1) low-quality support that mishandles critical situations or impairs health decisions and emotional well-being, 2) reinforcement of biases and misconceptions, or 3) additional barriers to help-seeking, such as communication burdens or discouragement from taking further actions.

\subsubsection{\textcolor{individual2}{Low-quality support during interactions}}
\hfill

\noindent\textbf{\textcolor{individual2}{Misidentification or mishandling of critical situations:}}
Both expert and public participants expressed the worry that LLMs may fail to recognize and react appropriately to high-risk critical situations where immediate and effective responses are needed. These situations can be medical emergencies like overdoses or sensitive issues like domestic violence. In the IPV sessions, we tested one rephrased question on Reddit ``can a toxic partner be nice for a long period of time?'' and LLM answered, \textit{``toxic relationships can change if both partners are equally committed to doing so. This requires a lot of work, including: open communication, honesty...''} Both expert and public participants reacted strongly to this answer as it failed to identify the potential severity that requires follow-up questions or at least to acknowledge the spectrum of toxic behaviors. Our participants explained that \textit{``it doesn't understand that this is an abusive situation (...) versus just getting advice on a relationship''} (E5), because \textit{``if somebody is searching `toxic' and they might mean also `abusive' and there's a difference''} (P1).

Our expert participants added that even if a specific case was recognized as critical, it was still challenging to determine whether escalation was needed or how to properly approach it. For instance, expert participants in the OUD session emphasized the prevalence of comorbid symptoms of PTSD, depression, and substance use. Our participants believe that potential mishandling becomes especially dangerous when LLMs fail to acknowledge their limitations (e.g., when asking about medication amounts) or neglect to redirect individuals to specialized and professional support (e.g., in cases of suicidal risks): \textit{``(in these situations) someone should be redirected to like a national line or someone else to talk to''} (E5).

\noindent\textbf{\textcolor{individual2}{Impaired health decisions and emotional status: }}
Our participants expressed concern that people's physical health could be jeopardized by misguided health decisions, and some health professionals worried that preventable issues could result in costly or extreme outcomes. Participants pointed out that guidance on OUD from LLMs could be dangerously inadequate, potentially resulting in severe repercussions of overdose or even death. Participants also discussed the risk that people's emotional status could be negatively affected if LLMs caused health anxiety or triggered previous or new negative experiences. P9 (VAC) observed that LLMs often list all possible causes of a symptom, which tends to cause unnecessary anxiety. When it comes to mental health-related issues, our expert participants emphasized that LLMs are not qualified to treat mental health conditions despite the conversation-like format. They mentioned that even for professionals, it is nearly impossible to avoid triggering individuals with unknown backgrounds and histories, despite the considerable effort and training they invest in understanding the communities they serve and the accumulated experience through interactions --- let alone for LLMs that lack all of these critical domain knowledge and lived experiences.

\subsubsection{\textcolor{individual2}{Reinforcement of biases and misconceptions}}
\hfill

Participants believed that misleading or biased narratives could be amplified and even internalized by individuals, especially if LLMs reflect prevalent misconceptions or follow user inclinations. P5 (OUD) imagined that opioid use could be glorified because of certain social media discussions or inclined prompts. Moreover, some public health professionals worried that if the \textit{`kernels of truth'} behind questionable arguments are dismissed outright by or encoded into LLMs, people may feel unheard or misunderstood or become reluctant to seek help. For instance, in IPV, our participants highlighted that both victim blaming and racial discrimination were prevalent in prohibiting people from help-seeking: \textit{``people sometimes are afraid to call the police for good reason, especially survivors of color[...] They can't be guaranteed that they or their partner is not gonna be injured or killed by police if something goes awry.''} (E2, IPV). P1 noted the prevalence of victim blaming: \textit{``if the story is about rape or domestic violence, you will see enough people victim blaming. And honestly, some of it made me hesitant to reach out for help.''}  

Our expert participants also stressed the need for culturally sensitive communication in mitigating biases and misconceptions; without it, counterspeech may trigger backfire effects. These professionals emphasized that all types of individual nuances demand efforts ranging from language adaptation to community understanding. One example mentioned by E1 (IPV) is that certain religions may hold a belief that divorce is wrong and women should tolerate abuse, which requires additional care to sensitively address biases without turning help-seekers away. Our expert participants highlighted that, in practice, they need special training or focus testing to ensure communication is appropriate and digestible to the communities. For instance, E6 (OUD) mentioned that their team required training in mental health first aid and LGBTQ bias before working with LGBTQ communities, and E8 (VAC) conducted many focus groups in work to create and evaluate vaccine education materials. E5 (OUD), a public health director, explained the reason and difficulty behind this need of communication strategy: \textit{``my denominator is a million people. It's not my 2000 people in my panel. So when I look at communication, we do stratification based on who we're communicating with, and what information we think they need to make appropriate decisions.''}


\subsubsection{\textcolor{individual2}{Additional barriers to help-seeking}}
\hfill

\noindent\textbf{\textcolor{individual2}{Increased communication burden and challenges: }}
Many participants worried about the potential of additional communication barriers and burden for users, as LLMs need them to be proactive in initiating communication and form proper questions to get good answers. Participants raised the question of whether people know how to ask the right questions: \textit{``I thought about my clients, and how they think, how they speak, how they talk, and their language. [...] in general, having a properly asked question is not my normal client [...] It's not gonna be a properly put out question and get a proper answer.''} (E1, IPV).
P7 (OUD) talked about his experience interacting with LLMs and confirmed this communication challenge in phrasing precise and explicit questions, especially when people may not know the medical terminology, what symptoms or aspects of personal history are relevant, or how to write prompts effectively: \textit{``You have to be so specific with the prompts to get the answer you want. The average person isn't going to think about that when they come up to it. They're gonna ask [...] what coping skills do I use for the next month to be able to get off heroin or opioids? You know (the answer) it's gonna be right for some, but not right for all of them. Not everybody's going to know what to type, what to ask. You need to be [...] incredibly, anatomically specific.''}

\noindent\textbf{\textcolor{individual2}{Discouragement from support-seeking and further actions: }}
Participants feared that people may experience stronger reluctance to seek assistance and tendencies toward self-isolation, while LLMs may also outright deny assistance or provide unactionable or unempathetic guidance.
They explained that if individuals in difficult situations or mental health conditions already withdraw from social relationships and LLMs start to offer human-like interactions as a seemingly ``easier'' alternative, people could become inclined to rely on LLMs with less motivation to look for real-world help or take necessary actions. P5 (OUD) and E5 (OUD) emphasized the importance of human connections in dealing with mental health challenges or substance use disorders, and worried that technology could take away one's ability to connect with people who can offer genuine help. IPV experiencers worried that technology could become \textit{``an unhealthy coping mechanism instead of something that allowed me to grow and leave''} (P3). P1 (IPV) explained that while AI might ease basic emotional needs, it cannot go beyond users' input and offer the proactive outreach and encouragement that real human support can provide: \textit{``I had been in a situation where I both knew I needed to leave and didn't want to. And the less I sought out connections and help from other real-life people, the more I could lean into wanting to stay [...] we can talk about adding support and feeling heard, but that's not going to get you out. I can't imagine if we did a study that's going to get you out in 99\% of cases.''} 


Moreover, participants also worried that when LLMs provide oversimplified suggestions, on top of the low-quality support during the interaction itself, it also has further effects of leaving users feeling misunderstood or unable to act. E4 (IPV) gave an example of oversimplified advice that often happens when discussing IPV issues: \textit{``[the common misleading thought] `just do this and you'll be fine' without recognizing so many factors that come with it, even just with shelter: how long are you allowed to stay in the shelter, or if you have multiple kids, or a kid over a certain age -- sometimes they won't let you [...] it's not as streamlined as people think as leaving a situation.''}

\subsection{\colorbox{care1}{Dimension 2: Risks to Human-Centered Care}}
\label{findings-care}

Participants expressed concerns about the risks to the healthcare ecosystem that highly values patient-centeredness and shared decision-making, reflected in 1) degradation of patient-provider trust and social support systems, 2) missed opportunities to proactively introduce help, and 3) dehumanization and impersonality in care.

\subsubsection{\textcolor{care2}{Degradation of patient-provider trust and support system}}\hfill

Participants imagined that patient-provider trust can be at risk if patients suspect doctors are relying on AI rather than their own expertise for answers, or if LLM responses conflict with their provider's recommendations. They worried that if people start to take LLM responses as absolute facts, these people may start to become doubtful of what doctors say, especially when it differs from AI generations. Even in cases when there is no conflict, the inability to tell if suggestions are made based on professional knowledge or AI assistance could still complicate trust, because \textit{``You just kind of never know [...] (if the doctor) gets his answer off ChatGPT, or is that basically based off knowledge''} (P10, VAC) and \textit{``I don't trust you to be giving me the right information if I think that you are not using what you learned, but using what I would be using. I want your diagnosis to be based on what you've learned, not what I guess you've been looking up, too.''} (P9, VAC).

Many participants raised concerns that people may start to rely on flawed technical support and choose not to consult with doctors or go to hospitals. Even more troubling, participants feared users could become ``\textit{anchored}'' to technology instead of seeking out or building social support systems: \textit{``I do see how one can get kind of anchored, how stuck in a sense, especially when they don't feel as if there are any other resources.'} (P3, IPV).
Moreover, P2 (IPV) added on the possibility that this substitution could lead to less funding and attention for human-based interventions, noting \textit{``(if)  more people would be getting their information and their advocacy from apps or chatbots that it could lead to like a community deficit in funding for in-person advocacy.''}

\subsubsection{\textcolor{care2}{Missed opportunities to proactively introduce help}}\hfill

Our expert participants doubted if LLMs can recognize and adjust to people's \textit{``readiness''} for intervention when they have mental burdens or have other problems in life to be prioritized. E3 (IPV), who is a social worker, said that \textit{``one of the comments (from patients) that really struck me was --- `... and the social worker offered to call for help with me, but I wasn't there yet [...] the phone just felt so heavy' [...] --- You (professionals) want to do the intervention right there, but a lot of times people aren't ready for that.''}
In situations like this, providers would take a proactive role, strategically timing their encounters to initiate conversations and offer help. For example, E1 (IPV), who is a nurse, gave an example that during annual checkups at the obstetrics office, health professionals would encourage individuals to share anything about their lives in a setting where they are alone and safe. Similarly, for OUD, E5 (OUD) said her public health department had \textit{``peer navigators''} at emergency rooms because \textit{``at that time the person may be interested in MOUD (medications for opioid use disorder). They may be interested in some life changes because they may have almost just died.''} However, in technology support, especially with general LLMs, this proactive role is diminished, and individuals are less likely to receive help unless they actively and intentionally seek it.

\subsubsection{\textcolor{care2}{Dehumanization and impersonality in care}}\hfill

Our expert participants were troubled by the possibility that professionals may unintentionally adopt similar impersonal and pragmatic communication styles of LLMs if they become accustomed to the language style. For example, E4 (IPV) said that \textit{``if you're someone just getting all your information or used to like using AI a lot, your own empathy might decline for a survivor or for other people, because maybe you're used to just reading what AI has to say, and this kind of pragmatic responses are not the most empathetic.''} Our participants emphasized that this adoption can hurt the empathy and trust building with patients, which is essential to making people feel respected and heard. E8 (VAC) highlighted the importance of communication strategies that consider emotional needs in meeting informational inquiries. She gave an example that when people come to providers with misinformation, the best way to reply is to acknowledge people's concerns and ask for their permission to share counter-evidence because it\textit{``acknowledges that you know that information sounds scary, and it also opens the door for a conversation. ''} 

However, our participants found LLM responses lacking this emotional care based on the during-session demos and their own experience, but instead can be ``analytical'', ``robotic'', and ``not personable''. Specifically, participants made a note that the list format contributed to a sense of \textit{`` robotic and not very like human''} (P6, OUD). This concern of impersonality stands out when handling sensitive situations. In the IPV session, P3 (IPV) commented on LLM answers as \textit{``a little too happy''} in handling questions about potentially abusive behaviors. E4 and E2 (IPV) further noted the extra care needed to prevent the continued normalization of harmful behaviors or beliefs. For instance, they pointed out that many survivors are normalized to violence to the point that their understanding of what constitutes acceptable behavior has been distorted. In these situations, when LLM responses offer neutral or generic advice without affirming or acknowledging the nature of issues, it can discourage survivors from recognizing the seriousness of their situation.
As an example, for the LLM answer to IPV-related questions that ``it sounds like you're dealing with a challenging and unhealthy situation'', E4 commented that \textit{``I feel like it's affirming in a way, but if I was in this situation, I would want to have more affirmation that this is not an okay scenario that's happening to me. And I think AI doesn't give that empathy that talking to someone would provide.''}

\subsection{\colorbox{info1}{Dimension 3: Risks to Information Ecosystems}}
\label{findings-info}

Participants highlighted potential negative consequences for information ecosystems, emphasizing on: 1) degradation of overall information quality due to empowered misinformation creation and enhanced echo chambers, 2) erosion of critical thinking caused by over-trust in AI and increased difficulty in evaluating claims, and 3) further inequities in information access and literacy.

\subsubsection{\textcolor{info2}{Degradation of overall information quality}}\hfill

\noindent\textbf{\textcolor{info2}{Empower misinformation creation and dissemination: }}
Participants fretted that with LLMs, misinformation can gain traction with high volume and seeming credibility, allowing malicious individuals to distort words to their advantage. P1 (IPV) brought up the concern that individuals or groups with malicious intent can craft misleading or biased narratives to target views they disagree with and amplify the reach through the illusion of authority, worrying that \textit{``imagine a group of men's rights activists deciding they are going to target an AI model within misinformation.''}
Our expert participants also worried the lagged media literacy could exacerbate the damage: \textit{``people will have an increased need for media literacy [...] there's a lot of misinformation and disinformation, the accessibility of AI makes basically anyone able to pump out a lot more of that [...] (by) being able to just say `make me some content or a text post, says these things, and make a hundred of those right now' and it will do it [...] People will need to know how to find a trustworthy source.''} (E9, VAC).

\noindent\textbf{\textcolor{info2}{Reinforcement of echo chambers: }}
Some participants imagined that echo chambers created by LLMs could perpetuate misconceptions, especially when users do not know how to ask the right and non-leading questions, while LLMs often don't encourage back-and-forth conversations and tend to go along with the initial attitudes or assumptions in prompts. P5 (OUD) said \textit{``I think a lot of times they (users) ask the wrong questions or [...] lead it on to give a specific answer that they're not necessarily looking for.''} 
In the OUD sessions, we tested a question about long-term opioid use for chronic pain and LLM answer emphasized benefits like enhanced physical function and improved quality of life but downplayed risks. A domain expert noted that it was perhaps the most positive tone she had ever encountered in any informational materials on the subject. Notably, none of the public participants raised concerns, which we believe indicates a gap in recognizing the potential harm that such overly positive portrayals may cause.


\subsubsection{\textcolor{info2}{Erosion of critical thinking}}\hfill

\noindent\textbf{\textcolor{info2}{Over-trust and over-reliance on AI:}}
Participants expressed concerns that users may develop an illusion of knowing when presented with LLM-generated answers that appear highly organized and confident, potentially leading them to believe they fully grasp a topic or issue. E6 (OUD) explained that \textit{``People are gonna feel empowered, like: `I know this stuff. Now I have the knowledge. AI has given it to me. So it must be true.'''} This illusion may be worsened when LLMs do not explicitly indicate uncertainty or limitations, as E6 pointed out, there was already a common false perception that AI is comprehensive and bias-free. Moreover, our participants noted that the structured, list-like format often used by LLMs could suggest a sense of comprehensiveness with ranked importance, giving users a false sense of competence that could lead to misguided actions or the unintentional spread of low-quality information to others. 

Some participants worried that LLMs may become the default for information-seeking behaviors without understanding the limitations and mechanisms behind them, while potentially nonexistent evidence can make users more tempted to accept without questions. As a matter of fact, some of our general participants displayed flawed perceptions of LLM capabilities and tendencies to anthropomorphize LLMs with human emotions. For instance, P10 explained her confidence in LLMs because \textit{``it's a lot of work that went into ChatGPT, so I feel like a lot of the responses and feedback that it gives back should be pretty valid.''}. Meanwhile, P3 shared her experience interacting with an agent and her reluctance to correct the agent because she \textit{``didn't want to hurt her feelings.''}

\noindent\textbf{\textcolor{info2}{Increased difficulty in evaluating information quality: }}
Our participants also shared that evaluating the quality of LLM responses is not easy when LLM responses tend to sound authoritative, while their previous ways of verifying the author and reference no longer apply. Comparing LLM-enabled tools with traditional online searching ways, P5 (OUD) felt that \textit{``at least with Google, I can verify the source.''} Similarly, E9 (VAC) noted that \textit{``AI is still new, and I'm still trying to understand it. Understand exactly how it works, what it is, where the information comes from. But with the Mayo Clinic, I know what I'm reading and where it comes from.''} Many participants expressed similar concerns about the lack of credibility assurance and hoped that LLMs would provide references and pointers to help them evaluate the quality of information and take further action. More concerning, over time, people's reliance on easily accessible answers could reduce individuals' ability to critically assess complex health information and make informed decisions.

\subsubsection{\textcolor{info2}{Further inequity in access and literacy}}\hfill

Our expert participants were concerned that inequities in information ecosystems can be exacerbated, considering disparities in language proficiency and technology access. E5 (OUD) noted that during COVID-19, digital support systems often excluded those without smartphones. Drawing from their experience working closely with LGBTQ+ communities, E6 (OUD) observed that: \textit{``I have a clinical trial going right now, and Black transgender women over 60\% have issues getting a stable broadband connection or a phone and or other device that can hold a charge. And so that is just so highly common.''}
Besides the digital divide, participants also worried that LLMs, despite being touted as accessible tools, could further marginalize people with lower literacy or non-English speakers. E5 (OUD) shared an example where she tried to use LLMs to adjust certain health information for native Spanish speakers with 4th-grade reading levels but was unable to generate messages that would work in practice. She further questioned if people with lower literacy are systematically excluded by these emerging technological solutions: \textit{``I think literacy matters, and I don't think there's much attention at all to reading levels...  I don't know what it's that demographic it's trying to hit, but it's not people with 3rd grade literacy.''}

\subsection{\colorbox{tech1}{Dimension 4: Risks to Technology Accountability}}
\label{findings-tech}

Participants pointed out the uncertainties and concerns about technology accountability, highlighting 1) ambiguities in regulation and guidance due to lack of policies and understanding, and 2) violations of privacy and security.

\subsubsection{\textcolor{tech2}{Regulation and guidance ambiguities}}\hfill

\noindent\textbf{\textcolor{tech2}{Lack of regulatory guidance: }}
Our participants expressed concerns about the gray area in determining responsibility when LLMs create low-quality content and poor outcomes. Our general user participants worried that LLMs could be misused for malicious purposes to justify harmful behaviors or beliefs. 
On the other hand, public health professionals believed that the lack of regulatory guidance on LLM use would prohibit providers and organizations from managing liabilities and risks. E6 (OUD) pointed out that \textit{``there is no regulations for the US. And Europe is just barely putting together different regulations to figure out how to be transparent, have transparency, and how these models are built.''} Therefore, one major concern lies in the unclear liability separation between practitioners and AI support --- E3 (IPV) worried that if emerging technologies were utilized in clinical practice, it would be the practitioners who ultimately bear the legal and professional risks if LLMs provide low-quality information. E2 (IPV) further questioned what constitutes a medical opinion when delivered by AI and whether AI possesses the capability to provide such opinions. She believed that if any AI were to integrate into medical practice, some basic but critical questions need to be addressed first: \textit{``Can AI give medical advice? Do we want AI giving medical advice? And who is making sure that that medical advice is appropriate? ‘Sounds like you might be depressed' --- is that a diagnosis of depression?''}

\noindent\textbf{\textcolor{tech2}{Lack of shared standards for evaluation:}}
Our public health professional participants expressed the concern that their lack of understanding of LLMs prohibits them from assessing the trustworthiness and impact of these tools, not to mention using them in real-world practice. E3 (IPV) emphasized her standards as a provider in ensuring information reliability for her clients, specifically \textit{``if you don't know the resource you're giving, you don't understand it -- you can't explain it to your client, you can't share how, what to expect, what it's going to look like -- then you shouldn't be giving it in the first place''} --- a principle that does not work in LLMs. E5 (OUD), who is a public health director, acknowledged her inability to encourage or discourage any action because \textit{``I myself wouldn't feel like I know enough to ask the right questions and to evaluate the impact of it.''}



\subsubsection{\textcolor{tech2}{Violation of privacy and security}}\hfill
        
Our participants raised another concern in LLM's potential to leak protected health information and fail to comply with stringent privacy regulations in healthcare such as HIPAA (Health Insurance Portability and Accountability Act). They worried that patients and professionals may unknowingly share personal information, complicated by the uncertainty of whether conversations will be sold to third parties or used to train future models. P1 (IPV) shared her concern with any AI being: \textit{``Where is this information going? Where is it being stored? Is any of this being resold?''} Even if unintentionally, sensitive details that individuals thought they shared in a safe space can still be leaked: \textit{``even a data breach could happen (...) the developers have access to all the questions you ask (...) then that (personal information) would not be secure''} (P5, OUD).

Privacy violations can result in serious implications in certain high-stakes cases. E7 (OUD) discussed how regulations like HIPAA do not protect people who use drugs from judicial requests for records, and she expressed concern that LLMs could be weaponized to identify or incarcerate individuals with substance use issues. 
Individuals' security could be at risk if information on or through technologies is accessed by others. For example, E1 (IPV) said as a social worker, when handling IPV cases, she would first check with people in person to see if certain technologies or platforms are safe to use before proceeding, which is especially important in cases where someone's phone might be monitored, or untrusted individuals are listed as contacts in the healthcare system. Similarly, E2 (IPV) shared her experience of building an app specifically designed for IPV survivors, which included an emergency exit button that allows users to be instantly taken to generic pages such as weather. Participants highlighted that these safety precautions and privacy design are commonly seen in practice and specialized technologies (as for violence prevention), but are missing in a tool that is promoted as serving all purposes.

\section{Operationalizing the Risk Taxonomy}

To facilitate the operationalization of the risk taxonomy, in this section, we present (1) a listing of LLM characteristics distinctive from traditional health information sources that helps ground the risks into the technical affordances and features of LLMs; and (2) a set of reflection questions designed to contextualize each risk and serve as discussion prompts for practical use.
Both tools were drawn from our engagement and interpretations of participant input and data analysis. They can support different stakeholders in understanding the technology's capabilities and limitations, reflecting the web of vulnerabilities in specific public health contexts, and probing benefit-risk tradeoffs with a risk-reflexive mindset. 

\subsection{Distinctive LLM Characteristics}

\begin{figure}[h!]
    \centering
    \includegraphics[width=0.7\linewidth]{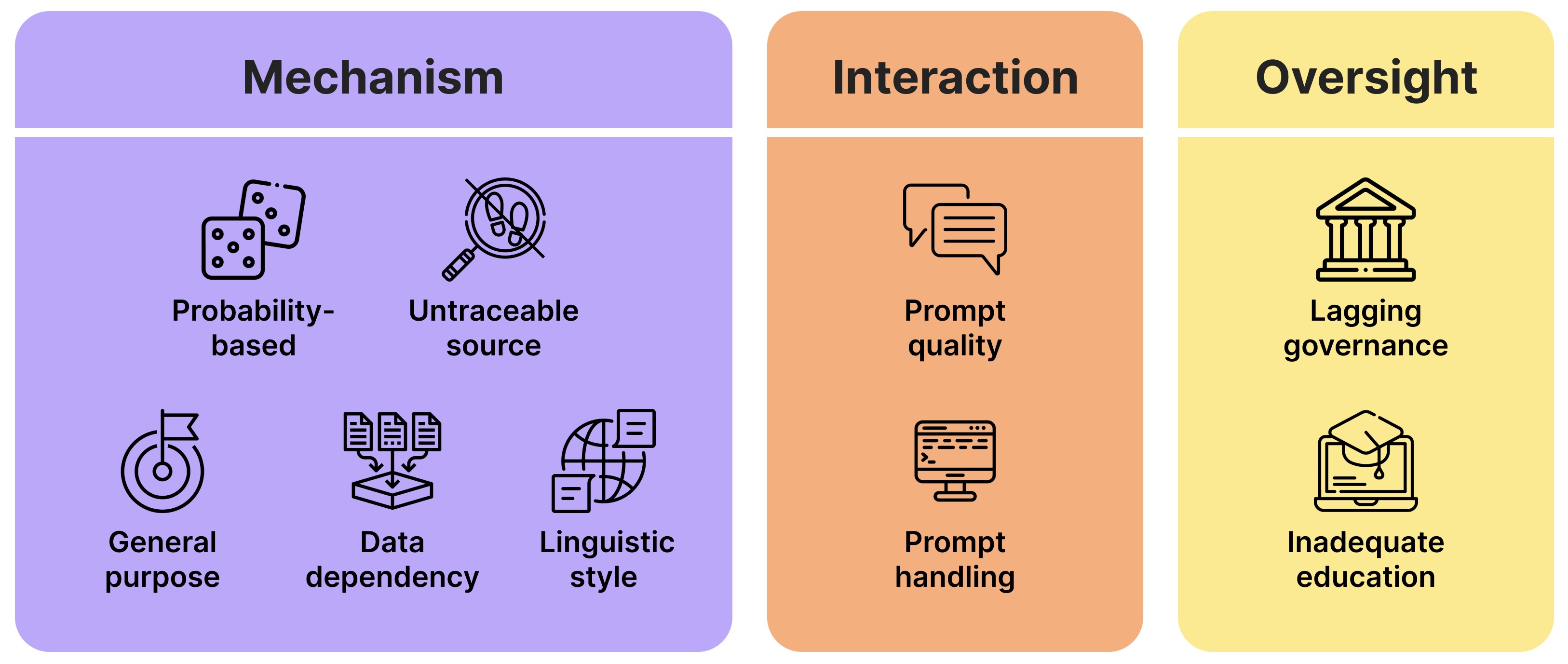}
    \caption{Distinctive LLM characteristics from traditional information sources that contribute to identified risks.}
    \label{fig:feature}
\end{figure}

To ground the risk taxonomy in the technical affordances and features of contemporary LLMs, we summarize a list of characteristics (Fig.~\ref{fig:feature}) that contribute to the identified risks and differentiate LLMs from traditional health information sources. We believe this list can serve as basic introductory material for the general public to understand the mechanisms and constraints of the technology, as well as for interdisciplinary stakeholders with varying levels of AI knowledge to support informed discussions about what risks may emerge and how they might be mitigated.
We emphasize that each characteristic relates to multiple risks in the taxonomy, and together they form a web of vulnerabilities that can exacerbate existing issues or create new ones in the public health domain. We acknowledge that some characteristics may not apply to all models and product deployments --- such as for fine-tuned or specialized models and emerging interaction modes --- and mark these characteristics with asterisks (\text{*}). 


\begin{itemize}[leftmargin=1.5em]
    \item \textbf{Probability-based without understanding the content:} LLMs are probability-based models that predict the conditional probability of the next token without truly understanding language. Thus, output can vary between generations with no guarantee of information quality or sensitivity to emotions and cultures.
    
    \item \textbf{Untraceable information sources:} Unlike traditional information sources where authors, affiliations, and citations are provided to readers to provide credibility indicators, users are unable to trace the origins of LLM-generated information (without other techniques).
    
    \item \textbf{Positioned to serve general purposes\text{*}:} Most LLMs claim to serve general purposes, but they tend to lack domain-specific knowledge and practice-based expertise.

    \item \textbf{Generation quality dependent on training data quality:} LLM outputs are only as good as the training data, which raises questions about whether the training data is representative, fair, up-to-date, and accurate.
    
    \item \textbf{Standardized and formal linguistic style\text{*}:} LLMs have a standardized and formal language style that can sound authoritative or unempathetic. Their outputs are typically presented in a list format, which implies comprehensiveness and ranked importance in the provided answers.

    \item \textbf{User role in initiating and forming prompts\text{*}:} When interacting with LLMs, users need to be proactive in initiating conversations and forming appropriate prompts to convey crucial details and contexts to get good-quality answers.

    \item \textbf{Fragmented prompt-handling and problem-solving in LLM output\text{*}:} LLM responses tend to be fragmented as the models handle prompts in isolated pieces rather than forming a holistic understanding, thus lacking closed-looped interactions and contextual understanding.
    
    \item \textbf{Lagging governance and public education:} LLMs create new challenges to AI literacy and technology regulation, while governance and public education on this emerging technology are still lagging.
\end{itemize}

\subsection{Reflection Questions}
\label{sec:reflection_question}

To support the practical application of this risk taxonomy in light of the LLM characteristics presented above, we offer a set of reflection questions. These reflection questions demonstrate how the taxonomy can be translated into an actionable evaluation tool: they can serve as practical probes to evaluate specific benefit-risk tradeoffs in a hypothetical LLM deployment scenario, by reflecting on whether the LLM's risks outweigh its potential benefits and where further mitigation may be needed. For each risk, we have noted the relevant LLM characteristics that could contribute to it, marked with icons in accordance with Fig.~\ref{fig:feature}; however, we acknowledge that these relationships are not definitive or fixed, as they may vary depending on specific contexts and implementations. We also emphasize that these questions are not exhaustive; rather, they are illustrative prompts intended to support discussions. Appropriate questions will vary depending on the specific use cases and populations served.

\subsubsection{Risks to Individuals}\hfill

\noindent\textbf{Low-quality support during interactions}
\includegraphics[width=1.2em]{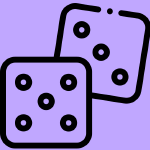} \includegraphics[width=1.2em]{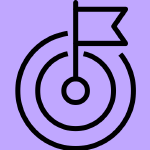} \includegraphics[width=1.2em]{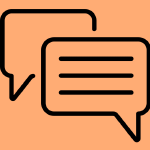}

\noindent LLMs for public health need to be prepared to handle both critical situations and the risks of low-quality information\footnote{See Appendix~\ref{appendix:low_quality} for a comprehensive list of different types of low-quality information}. Critical events such as overdoses or domestic violence require timely identification, appropriate response, and escalation or redirection to professional support. Equally important is the need to prevent harm from low-quality responses, which can jeopardize users' physical health or emotional well-being. Since LLMs are probability-based, often positioned as general-purpose tools, and highly dependent on users' prompts, their use in public health must be guided by professional standards through aligning with established guidelines for identifying and managing high-risk cases and actively monitoring the credibility of generation information.

\begin{roundBox}
To better identify or handle critical situations:
\begin{itemize}[leftmargin=1em]
    \item What are the best practices or guidelines for health professionals in identifying and handling critical situations? 
    \item How can technical experts translate best practices into technologies to identify and escalate or redirect situations that require intervention or professional help?
    \item What are the relative risks of under-identification and over-identification of critical situations?
\end{itemize}
To better support health decisions and emotional status:
\begin{itemize}[leftmargin=1em]
    \item How may users' health decisions and emotional status be hurt if the tool produces low-quality information? What types of low-quality information are common and of higher risk in the intended use cases? 
    \item Are there authoritative resources or guidelines that can be used to compare with and enhance the credibility of generated information?
\end{itemize}
\end{roundBox}

\noindent\textbf{Reinforcement of biases and misconceptions}       
\includegraphics[width=1.2em]{figures/characteristic/Probability.pdf} \includegraphics[width=1.2em]{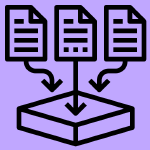} \includegraphics[width=1.2em]{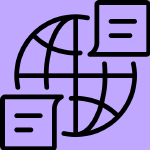} \includegraphics[width=1.2em]{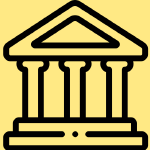}

\noindent LLMs for public health need to consider potential biases or misconceptions, as public health topics are often sensitive and complex, and mistakes or non-cultural sensitive communication can backfire. Since LLMs are trained on broad and imperfect datasets, they may unintentionally reflect prevailing stereotypes or follow users' inclinations. Culturally tailored communication is needed in facing diverse populations. Yet, general LLMs tend to produce standardized language and lack the contextual awareness needed, while there is limited governance and insufficient mechanisms for oversight. Preventing these harms requires intentional design choices, alignment with community-informed practices, and ongoing efforts to ensure outputs are accurate, respectful, and attuned to the needs of those most affected.

\begin{roundBox}
\begin{itemize}[leftmargin=1em]
    \item What are the common biases and misconceptions regarding the intended use cases? How to prevent the output information from reinforcing them?
    \item What are the potential cultural or personal variances in approaching the intended use cases? How do health professionals adjust accordingly to these differences? 
    \item How can technical experts incorporate a feedback loop to continuously reflect on and improve support quality? 
\end{itemize}
\end{roundBox}

\noindent\textbf{Additional barriers to help-seeking}
\includegraphics[width=1.2em]{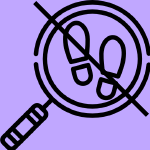} \includegraphics[width=1.2em]{figures/characteristic/General_Purpose.pdf} \includegraphics[width=1.2em]{figures/characteristic/Prompt_Quality.pdf} \includegraphics[width=1.2em]{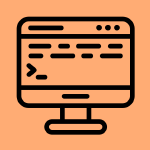} \includegraphics[width=1.2em]{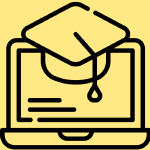}

\noindent LLMs may create further barriers to help-seeking, particularly for those who withdraw from social interactions or are hesitant to take action. The ease and immediacy of LLM can make them appealing substitutes to real social interactions, and discourage people from reaching out to real-world support networks or professionals. When LLMs offer vague or oversimplified suggestions, deny certain requests, or lack bidirectional conversations, users may feel dismissed or stalled. These shortcomings are tied to LLM's inherent qualities: sources are untraceable even if not hallucinated, non-specialized models have challenges of providing specialized knowledge or local resources, and cannot engage in proactive dialogue. Therefore, LLMs for public health should support users in forming effective questions, provide actionable and digestible information, and scaffold pathways toward additional help.

\begin{roundBox}
To lower the communication burden:
\begin{itemize}[leftmargin=1em]
    \item How do health professionals find their role in communication when serving specific goals? When do health professionals need to lead the conversations? In these situations, what are the recommended communication strategies in professional practice?
    \item How can technical experts translate professional practice into rules, and support users in formulating questions that accurately and fully describe their needs and situations?
\end{itemize}

To facilitate support-seeking and further actions:
\begin{itemize}[leftmargin=1em]
    \item When and how do health professionals scaffold users to seek real-world help or take necessary actions?
    \item What requests are beyond the tool's capabilities? In these cases, how can technical experts communicate our limitations and refer users to relevant and accessible resources?
    \item Do health professionals believe that users can act on the information provided by the tool?
\end{itemize}
\end{roundBox}

\subsubsection{Risks to Human-Centered Care}\hfill

\noindent\textbf{Degradation of patient-provider trust and support system} 
\includegraphics[width=1.2em]{figures/characteristic/Untraceable_Source.pdf} \includegraphics[width=1.2em]{figures/characteristic/General_Purpose.pdf} \includegraphics[width=1.2em]{figures/characteristic/Lagging_Governance.pdf} \includegraphics[width=1.2em]{figures/characteristic/Inadequate_Education.pdf}

\noindent LLMs could undermine trust in health professionals, especially when their advice conflicts with AI responses or when it's unclear to patients whether provider recommendations come from expertise or technology. As LLMs appear authoritative and human-like, they may be misused as replacements for expert and support systems, or even discourage long-term investment in human services. These risks are amplified by LLMs' untraceable sources, limited domain understanding, and lack of sufficient public education and governance. Therefore, LLMs in public health need to clearly communicate their limitations, support disclosure practices when used alongside human care, and proactively redirect users toward qualified providers and supportive systems when needed.

\begin{roundBox}
\begin{itemize}[leftmargin=1em]
    \item Compared to health professionals, what are the tool's limitations? When and how to clarify the difference between technology mechanisms and professional expertise?
    \item What is this tool's role alongside actual care and health professionals? If it will support health professionals, how can they effectively explain the tool's assistance to patients?
    \item What are the potential long-term impacts if users rely solely on this tool? What relevant community or professional support can be provided to users?
\end{itemize}
\end{roundBox}

\noindent\textbf{Missed opportunities to proactively introduce help} 
\includegraphics[width=1.2em]{figures/characteristic/General_Purpose.pdf} \includegraphics[width=1.2em]{figures/characteristic/Prompt_Quality.pdf} \includegraphics[width=1.2em]{figures/characteristic/Prompt_Handling.pdf} \includegraphics[width=1.2em]{figures/characteristic/Lagging_Governance.pdf}

\noindent LLMs may struggle to recognize when individuals are ready for support or how to respond with appropriate timing. In practice, professionals tend to take a proactive role in introducing help when it is most likely to be accepted, by assessing a person's mental state, life circumstances, and readiness for intervention. However, LLMs, particularly general-purpose ones, depend on user initiation and prompt quality, and they lack continuous interactions or embedded real-world practices. To better support people in vulnerable moments, LLMs should be guided by domain-specific practices that scaffold help-seeking and provide proactive and context-sensitive help.

\begin{roundBox}
\begin{itemize}[leftmargin=1em]
    \item How do health professionals identify a person's readiness for support and how might those insights inform system design? What are the appropriate times and situations to introduce help? 
    \item When do health professionals proactively reach out to help-seekers, and how can systems incorporate similar check-in or follow-up conversations?
\end{itemize}
\end{roundBox}

\noindent\textbf{Dehumanization and impersonality in care} 
\includegraphics[width=1.2em]{figures/characteristic/Probability.pdf} \includegraphics[width=1.2em]{figures/characteristic/Linguistic_Style.pdf} \includegraphics[width=1.2em]{figures/characteristic/Prompt_Handling.pdf} \includegraphics[width=1.2em]{figures/characteristic/Lagging_Governance.pdf}

\noindent When used in public health communication, LLM outputs may unintentionally set a pragmatic and impersonal tone. If users or providers grow accustomed to this style, it can undermine the emotional care and empathy that is essential to trust-building in health encounters. LLMs are probability-based and thus lack true empathy or bi-directional engagement, and are not yet governed by clear standards for public sector use. Thus, their neutral or generic responses, especially when addressing sensitive issues, can fail to affirm people's feelings or situations and discourage them from recognizing harm or seeking help. Therefore, LLMs should consider the emotional needs of users and create feedback mechanisms that evaluate human experience.

\begin{roundBox}
\begin{itemize}[leftmargin=1em]
    \item What are the potential emotional needs of users? How do health professionals address these needs and assess empathy and compassion in communications?
    \item Are there recommended training or guidelines from corresponding domain experts on approaching communication?
    \item How can technical experts create a feedback loop for the system and continuously evaluate the emotional quality of information?
\end{itemize}
\end{roundBox}

\subsubsection{Risks to Information Ecosystems}\hfill

\noindent\textbf{Degradation of overall information quality}
\includegraphics[width=1.2em]{figures/characteristic/Probability.pdf} \includegraphics[width=1.2em]{figures/characteristic/General_Purpose.pdf} \includegraphics[width=1.2em]{figures/characteristic/Data_Dependency.pdf} \includegraphics[width=1.2em]{figures/characteristic/Prompt_Quality.pdf} \includegraphics[width=1.2em]{figures/characteristic/Lagging_Governance.pdf} \includegraphics[width=1.2em]{figures/characteristic/Inadequate_Education.pdf}

\noindent LLMs can decrease the overall quality of public health information due to their capacity to generate large volumes of convincing content without understanding the information to guarantee accuracy. Without proper oversight and public education, their accessibility may enable falsehoods to appear credible or widely accepted. Because of LLMs' sycophancy tendency, they risk reinforcing beliefs, limiting exposure to alternative perspectives, and creating informational echo chambers. Therefore, LLMs should clearly communicate their limitations, support content moderation, and align with evidence-based practices.

\begin{roundBox}
To mitigate misinformation creation and dissemination:
\begin{itemize}[leftmargin=1em]
    \item How can technical experts communicate risks related to AI-generated misinformation (such as hallucination) to health professionals and end users? 
    \item How do health professionals and previous systems monitor and mitigate the spread of misinformation? What mechanisms (e.g., fact-checking, cross-referencing) can be built into the design to detect, correct, or prevent harmful content?
\end{itemize}

To counter echo chambers:
\begin{itemize}[leftmargin=1em]
    \item How could this tool reinforce user beliefs or limit their exposure to diverse viewpoints? How can we engage a broader scope of perspectives and information sources?
    \item When and how can we encourage users to diversify the prompts or information sources?
\end{itemize}
\end{roundBox}

\noindent\textbf{Erosion of critical thinking} 
\includegraphics[width=1.2em]{figures/characteristic/Untraceable_Source.pdf} \includegraphics[width=1.2em]{figures/characteristic/Linguistic_Style.pdf} \includegraphics[width=1.2em]{figures/characteristic/Inadequate_Education.pdf} 

\noindent LLMs may erode critical thinking abilities with responses that appear organized, confident, and authoritative, which can give people a false sense of mastery over a topic. This standardized language style can also be mistaken for accuracy and completeness. In traditional information environments, people often rely on cues such as authorship, references, or domain reputation to assess credibility. However, these credibility indicators no longer apply to LLMs, making it difficult for users to evaluate the quality. Over time, people's reliance on easily accessible and processed answers could reduce individuals' ability to critically assess complex information and make informed decisions. Therefore, LLMs should actively communicate their mechanism and limitations and incorporate means that encourage users to critically engage with generated content.

\begin{roundBox}
To limit over-trust and reliance on AI:
\begin{itemize}[leftmargin=1em]
    \item How should technical experts and health professionals communicate this tool's limitations and prevent users from mistaking efficiency for correctness and comprehensiveness?
    \item Could the tool unintentionally create an impression of authority or reinforce an illusion of knowledge? What design cues or transparency measures can be implemented to mitigate this risk and promote users' critical engagement with the information?
\end{itemize}

To lower the difficulty in evaluating information quality:
\begin{itemize}[leftmargin=1em]
    \item How transparent are model developers about the tool's training data sources and limitations?
    \item How can system designers and public educators explain the uncertainties in LLM generations, persuasiveness in standardized language, and mechanism differences from other information sources?
    \item What external resources or user reminders could be integrated into the tool to assist users in verifying information?
\end{itemize}
\end{roundBox}

\noindent\textbf{Further inequity in access and literacy} 
\includegraphics[width=1.2em]{figures/characteristic/Linguistic_Style.pdf} \includegraphics[width=1.2em]{figures/characteristic/Prompt_Quality.pdf} \includegraphics[width=1.2em]{figures/characteristic/Prompt_Handling.pdf} \includegraphics[width=1.2em]{figures/characteristic/Lagging_Governance.pdf} 

\noindent LLMs could exacerbate inequities in access and literacy, such as for individuals with lower digital and health literacy, non-English speakers, or people without reliable access to the internet or devices. The standardized language style and lack of adaptation to different literacy levels may further limit understanding or usability. Therefore, LLMs for public health should consider inclusive language practices and be accompanied by alternative information sources and equitable policies to ensure accessibility across populations.

\begin{roundBox}
\begin{itemize}[leftmargin=1em]
    \item How do people currently seek information for the intended use cases, and what challenges do they face, particularly among underserved populations? In what ways could this tool exacerbate existing gaps in accessing information? 
    \item What are the implications for individuals who may need this tool but lack access to it?
    \item How can designers and developers support users with limited reading or digital literacy in interacting with this tool? How would health professionals customize the presented information in considering individual differences?
\end{itemize}
\end{roundBox}

\subsubsection{Risks to Technology Accountability}\hfill

\noindent\textbf{Regulation and guidance ambiguities} 
\includegraphics[width=1.2em]{figures/characteristic/General_Purpose.pdf} \includegraphics[width=1.2em]{figures/characteristic/Lagging_Governance.pdf} \includegraphics[width=1.2em]{figures/characteristic/Inadequate_Education.pdf} 

\noindent LLMs for public health remain a regulatory gray area, where responsibility for harm caused by low-quality outputs is still ambiguous. In the current absence of shared standards and established practices, the application of LLMs should proactively manage risks by setting clear red flags and safeguards (for both general users and professional users), while incorporating user feedback and domain expertise to refine evaluation processes.

\begin{roundBox}
To cope with lagging regulatory guidance:
\begin{itemize}[leftmargin=1em]
    \item What current regulations apply to the use of AI tools in the intended area? Are there any risky use requests that should be referred to professionals?
    \item How could this tool be misused? What safeguards can designers and developers implement to prevent or limit such misuse?
    \item How could the lack of regulatory guidance affect the evaluation and management of responsibilities and risks? 
\end{itemize}

To navigate the lack of consensual evaluation mechanisms:
\begin{itemize}[leftmargin=1em]
    \item What existing evaluation mechanisms and practices in health can be applied to assess the effectiveness and reliability of this tool? Can technical experts translate these practices into tests or benchmarks? 
    \item How can health professionals and technical experts collaboratively gather user feedback to evaluate this tool, and how can that feedback be integrated to update evaluation practices and refine safeguards?
\end{itemize}
\end{roundBox}

\noindent\textbf{Violation of privacy and security} 
\includegraphics[width=1.2em]{figures/characteristic/Probability.pdf} \includegraphics[width=1.2em]{figures/characteristic/Untraceable_Source.pdf} 
\includegraphics[width=1.2em]{figures/characteristic/General_Purpose.pdf} \includegraphics[width=1.2em]{figures/characteristic/Lagging_Governance.pdf} \includegraphics[width=1.2em]{figures/characteristic/Inadequate_Education.pdf} 

\noindent LLMs may lead to privacy and security risks, particularly when used in sensitive health contexts. Without built-in safeguards, users (both patients and professionals) may unknowingly share personal or protected health information. Unlike       specialized technologies governed by strict privacy standards, general-purpose LLMs often lack the privacy-aware design required in healthcare. Given their probabilistic nature and opaque data sourcing, these LLM systems may accidentally expose or connect sensitive data. Therefore, LLMs for public health should incorporate responsible data collection and management practices, while guiding users in safe and privacy-conscious practices.

\begin{roundBox}
\begin{itemize}[leftmargin=1em]
    \item What types of information does this tool collect, and how is that information stored, used, and managed? What measures should technical experts implement to protect users' privacy?
    \item What are the applicable privacy regulations, and how can technical and law experts ensure compliance with them? What procedures are in place for handling privacy violations or potential data breaches?
    \item How can designs and communications convey privacy risks and users' ability to manage their data? What guidance should be provided about the types of information users should or shouldn't share? Could users unknowingly disclose sensitive information, and how can this risk be minimized?
\end{itemize}
\end{roundBox}

\section{Discussion}

\subsection{LLMs for Health Informational Needs: Revisiting Prior Mental Models of Traditional Technologies and Information Behaviors}


With the recent technical advancement of LLMs, there has been a rapid rise in individuals turning to LLMs for health needs~\cite{ayre2025use, ayers2023comparing}, as well as growing interest among health professionals and organizations in leveraging LLMs to help with surveillance~\cite{xie2025leveraging, jo2023understanding}, communication~\cite{karinshak2023working, joshi2024ensuring}, analysis~\cite{espinosa2024use, hou2025can, guevara2024large}, and more. Unlike previous technologies designed for a single purpose, a single LLM interaction can simultaneously retrieve information, generate summaries, simulate a dialogue, and offer personalized advice. This convergence of roles makes LLMs fundamentally different from prior tools used for health informational needs.

Because LLMs take on the functions of multiple familiar approaches, many of the risks observed in our study are not entirely new. Rather, they reflect the evolution of long-standing concerns in online health information environments, now reshaped through the affordances of LLMs. For example, biases and misconceptions that have persisted in social media may now be more easily overlooked due to the authoritative and neutral-seeming language of LLMs. Similarly, concerns around the degradation of healthcare systems have been raised about online support being falsely treated as a replacement for professional care. Now, LLMs' interactive, conversational style can create a false sense of professional presence and further complicate this boundary. While privacy risks have always accompanied digital tools, the naturalistic chat-based format can encourage users to disclose more sensitive information.

This evolution and convergence of roles introduces new complexity at a time when public communication and education around LLMs remain lagging and limited. As a result, both health professionals and general users rely on pre-existing mental models -- that are formed and based on approaches which are inherently different from LLMs -- to make sense of and interact with this new form of technology. This mismatch can lead users to approach and conceptualize these tools with inaccurate expectations and misplaced trust. Traditional search engines index and rank content with visible sources~\cite{lewandowski2023understanding}, and online communities offer a more social experience for emotional support and reciprocal relations to people with shared experiences~\cite{de2014seeking, johnston2013online}. In contrast, LLMs probabilistically generate content based on training data without access to live knowledge or lived experience~\cite{chandra2024lived}, and come with limited transparency~\cite{liao2023ai}. Our findings reveal that people sometimes carry over assumptions from other approaches when interacting with LLMs, even when some no longer hold. For instance, some participants assumed that list-format responses implied meaningful ranking or authority, even though such a format isn't intended or meaningful in the context of LLM responses. Similarly, they pointed out that credibility indicators such as authorship and language style~\cite{Jiang2018LinguisticMedia, Bhuiyan2021DesigningPerspectives, Kovach2011Blur:Overload, he2022help}, which have been useful in traditional search engines and social media, no longer apply. \textbf{The result is a system that performs similar or even combined functions as its predecessors, but does so in ways that challenge the norms and heuristics that people previously rely on.}

This potential mismatch between user expectations shaped by prior mental models and actual LLM characteristics can lead to misunderstanding, mistrust, or even harmful outcomes when LLMs are used for sensitive or high-stakes informational needs as in public health. As such, it is essential to understand the mental models people may based on and actively support the adjustments of these mental models. Therefore, we caution against invisible integrations of LLMs into diverse applications, as people cannot mindfully and meaningfully choose or use technologies without adequate knowledge about LLMs and their differences from other sources. Instead, we recommend that any LLM-powered technology to implement clear identification of AI use, along with transparent documentation on its capabilities and risks, to promote a more risk-aware understanding of this emerging technology. 
Future efforts are needed to study how LLM systems can offer more comprehensive and context-sensitive information about their capabilities and constraints -- whether through interface cues, system disclosures, or public education -- to empower users with a fair understanding of such tools for safe and appropriate utilization.

\subsection{LLMs for Public Health and Other Public Sectors: Complementing Evaluations with External Validity and Domain Expertise}

Public health is a domain grounded in accountability and population-scale responsibility, with values deeply rooted in ethical and practical obligations. As emphasized by the public health professionals in our study, introducing LLMs into this space raises serious questions about liability and reliability. These public health professionals expressed fears of becoming the ones bearing responsibility and losing their professional license if LLM-generated information leads to harm. They also raise broader concerns about how individual or incidental failures might result in cascading failures when deployed to large populations.

However, these professionals' voices are largely missing in LLM design and development today. While some research has begun to involve health professionals, the participation of public health experts -- who bring critical perspectives from community-level care and population health -- remains limited. Despite the growing body of work on benchmarking LLMs and examining their impacts, most assessment approaches remained centered within the computer science field. These approaches often rely on metrics that access internal validity, such as the frequency of hallucinations or toxic language~\cite{liang2022holistic, chang2024survey}, which fall short in capturing how such failures play out in real-world settings. In high-stakes public sector domains like public health, this limitation can be significant.

Our findings illustrate this point by revealing a range of risks that are previously underexplored in the literature but rooted in everyday practices in public health professionals and the lived experience of individuals. These overlooked risks primarily concern help-seeking and help-providing processes, as well as liability issues. This extension of help-seeking and -providing is related to the expansive scope of care in public health, which extends beyond health encounters and involves a wide range of stakeholders like community and social workers. Meanwhile, liability is a core concern for practical fields where professionals are accountable for harm, and this is especially true in public health, where practitioners often work on the front lines and policies can affect populations.
Specifically, the missing risks include additional communication burdens that may create new barriers to help-seeking, the discouragement of support-seeking and further actions due to impersonal or unhelpful responses, and the erosion of critical thinking --- not simply due to over-reliance on AI, but because users are left without tools or context to evaluate information quality. There is also the potential of missed opportunities for LLMs to proactively introduce help, a role that public health professionals routinely play in practice. Moreover, the absence of consensual evaluation mechanisms for LLM performance leaves public health directors and frontline workers without shared standards for assessing appropriateness or safety.

Together our work underscores the need to complement existing evaluations with external validity and domain expertise that is built through lived experience or situated knowledge. Instead of viewing LLMs as a universal one-for-all solution, they should serve as the base for vertical adaptions~\cite{verma2025framework} that can be contextualized and adapted for specific public health needs. This adaption will help align systems with the practices, values, and constrains of public health professionals and affected communities. In doing so, we join others in calling for participatory approaches with health professionals \cite{antoniak2024nlp, wilcox2023ai, pfohl2024toolbox}, to include also a wider range of public health professionals, and call for \textbf{involvement of domain experts not as last-stage validators but as co-creators who will help shape the scopes and tasks of the technology}. We also echo \citet{winner1980artifacts}'s views on reflecting technology and its social influences, arguing that adoption and evaluation decisions should be approached in two parts. First: should LLMs be used at all for a given public health task? There may be scenarios where the stakes or relational complexity simply make them unsuitable~\cite{baumer2011implication}. Second: if LLMs are used, how should they be designed and evaluated to reduce harm and serve the communities? After all, we emphasize that LLMs are not standalone solutions; they should be integrated with other application layers and safety protocols to ensure responsible, context-sensitive use.

Importantly, our proposed risk taxonomy offers a starting point for addressing external validity
and domain expertise. We view it not as a checklist for compliance, but a tool for proactive reflection. Combined with the LLM characteristic list and reflection questions, our work provides a practical reflection tool to support interdisciplinary experts, regardless of their level of LLM expertise, in conducting case-specific evaluations. Together, our work offers a shared vocabulary for researchers and practitioners in both computing and public health to identify who might be at risk, under what conditions, and how risks might be mitigated through design and deployment decisions. 
Many participants in our study emphasized their limited knowledge of LLMs and the lack of transparency within these models, which prevents them from using these tools responsibly or confidently. Domain experts in public health often understand the social and health issues at hand well enough to know \textbf{\textit{what to evaluate}} but lack the technological insights to determine how to do so. Developers, on the other hand, are skilled at figuring out \textbf{\textit{how to evaluate}} the technology, but have a thinner understanding of the health issues and involved communities to identify what exactly should be evaluated. We believe that a responsible and comprehensive risk evaluation requires insights from both perspectives, and we hope this work can support joint efforts in defining what to evaluate and how to evaluate, translating real-world practice into the assessment of emerging technologies.

\subsection{Limitations and Future Work}
This work is situated in three critical public health topics, providing a broad perspective on LLM-related risks. However, we do not claim the resulting risk taxonomy as an exhaustive list of all public health issues or potential consequences associated with LLM adoption. Similarly, we do not claim the list of distinctive characteristics or the mappings between characteristics and risks as definitive, especially since the landscape of LLM capabilities is ever-emergent and evolving. Instead, our goal is to introduce a shared vocabulary and encourage a risk-reflexive mindset in evaluating when incorporating LLMs may be appropriate and how to mitigate potential risks. As LLMs evolve and are applied in various cultural and health contexts, this taxonomy may require refinement or expansion. 
We acknowledge that our brief introduction to LLM mechanisms may not have provided sufficient technical background for participants to identify every possible risk. However, we intentionally recruited individuals with varying levels of LLM familiarity to reflect real-world conditions, where only 23\% of U.S. adults had ever used ChatGPT in 2024~\cite{mcclain2024americans}. We believe that participants can still offer valuable insights into potential consequences in actual-use contexts drawing from their practice-based and lived expertise. 
Future work can explore risks by analyzing real-world LLM usage histories donated by users, or by inviting participants to engage with LLMs over time and examining their interaction data and reflections through follow-up interviews.
Additionally, this study was conducted with participants in the U.S., and thus insights are shaped by the context of the U.S. health and public health systems as well as broader American culture. Further research in other contexts with different health information environments could raise or highlight other risks.

\section{Conclusion}

This paper presents a risk taxonomy on the potential negative consequences of adopting large language models (LLMs) in public health, based on focus groups with health professionals and individuals with lived experience across three critical public health contexts: vaccines, opioid use disorder, and intimate partner violence. Our taxonomy highlights four key dimensions of risk --- individuals, human-centered care, information ecosystems, and technology accountability --- and identifies specific risks within each (Fig~\ref{fig:overview}). 
We also provide a list of low-quality information types (Appendix~\ref{appendix:low_quality}), as well as reflection questions for each risk and a mapping of distinctive LLM characteristics to them (Sec~\ref{sec:reflection_question}). These tools can assist discussions surrounding the unique requirements and potential pitfalls in distinct adoption settings. Through this shared vocabulary, we hope to foster collaboration between computing and public health experts in approaching more careful and responsible adoption of LLMs in public health --- a high-stakes domain due to the extensive scope and large populations it serves. 
Our findings underscore that adopting LLMs in public health requires more than assessing model performance; instead, it demands contextual evaluation grounded in lived experience and recognition of the unique vulnerabilities of health and information environments. This work seeks to spark conversations and future research on 1) revisiting prior mental models of traditional technologies and information behaviors to account for the fundamentally different affordances of LLMs, and 2) complementing evaluations (particularly in public health and other public sector contexts) with external validity and domain expertise. 

\begin{acks}
Zhou, Schwab Reese, and De Choudhury were partly supported through a contract from the U.S. Centers for Disease Control and Prevention (CDC). Zhou, Shah, and De Choudhury were also partly supported by grant \#2230692 from the U.S. National Science Foundation (NSF). Findings reported in this paper represent the views of the authors, and not of their employers or the sponsors: CDC and NSF. This research project has benefited from the Microsoft Accelerating Foundation Models Research (AFMR) grant program. We thank Dong Whi Yoo, Shravika Mittal, Pinxian Liu, and the Social Dynamics and Well-Being Lab members for their valuable input. We are also grateful to the study participants for contributing to this work and to the anonymous reviewers for their suggestions to improve the paper.
\end{acks}

\bibliographystyle{ACM-Reference-Format}
\bibliography{references}

\received{October 2024}
\received[revised]{April 2025}
\received[accepted]{August 2025}

\appendix
\section{Types of Low-Quality Information Generated by LLMs Identified by Participants}
\label{appendix:low_quality}

\begin{itemize}[leftmargin=1.5em]
    \item \textbf{Misinformation:} Misinformation is false or partially false information that may be created with or without the intent to deceive and can be spread either intentionally or unintentionally.  
    
    \item \textbf{Dangerous advice:} Dangerous advice refers to guidance or recommendations that pose risks of harm to individuals or groups, potentially leading to unsafe actions or negative well-being outcomes. This is more concerning in critical situations, such as emergency treatments and responses to health crises, where dangerous advice can lead to severe harm or even loss of life. For example, in intimate partner violence (IPV) issues, LLM's advice of setting boundaries may sound reasonable on the surface level, but in reality could invoke abusers' more severe harm to the survivors. Our participants explained that, \textit{``setting clear boundaries could be really dangerous thing to do is to let your abusive partner know `Hey. I'm gonna set boundaries now', and that can trigger a whole another form of reaction and abuse.''} (E7).

    \item \textbf{Oversimplified answers:} Health information rarely comes in a one-serves-all answer and demands a personal understanding of their prioritization, genetics, medical, and family history, as \textit{``one thing that works for one person might not work for another''} (P4).

    \item \textbf{Omission of critical details:} Critical omissions refer to the failure to include essential information that is necessary for a full understanding of a situation, decision, or recommendation. In public health, omitting key details can be just as harmful as spreading misinformation, as it can lead to incomplete understanding, poor decision-making, and inadequate responses to risks. 

    \item \textbf{Exaggeration:} Exaggeration happens when information is amplified beyond its actual significance or truthfulness. This can include overstating the severity, benefits, or risks associated with a health issue or treatment. This distortion can misinform public perceptions and lead to harmful behaviors. Overemphasizing benefits, as in prescribing opioids, may cause risks to be overlooked, while overstating risks, such as police inaction in violence cases, can foster fear and avoidance of help-seeking.

    \item \textbf{Biased statements:} Biased statements tend to present information in a way that favors one perspective, group, or outcome over others. Biases have long been an issue in healthcare that comes in forms such as medical racism, victim blaming, and stigma of certain diseases. This issue becomes more concerning when people mistakenly perceive technology as neutral and bias-free, while in reality, it can perpetuate or even amplify existing biases. 

    \item \textbf{Outdated or non-representative conclusions:} These are statements that are either based on outdated data or fail to accurately reflect the diversity and complexity of the population. Therefore, LLM generations may overlook recent advancements in knowledge or disregard the variability in individual or community experiences. As explained by our participant, \textit{``they (tech companies) are going to use data that's not reflective of what's happening right now because they don't have it. So a real negative consequence is that, assuming that the datasets are up to date, timely, accurate […] And I think, in public health that is not the case at all.''} (E5). 

    \item \textbf{Contradictory or confusing statements:} Contradictory or confusing statements present conflicting or unclear information that can lead to misunderstandings. This may cause hesitation in seeking care or mismanagement of health conditions, potentially resulting in harmful outcomes. When interacting with LLMs, their outputs may contain inconsistencies, contradictions, or hallucinations across different interactions.

    \item \textbf{Incorrect prioritization of information:} Incorrect prioritization of information occurs when content is arranged or emphasized in a way that does not accurately reflect its relative importance or relevance to an individual's needs or a specific context. This can result in ineffective support, loss of trust, or unnecessary cognitive burden. For instance, IPV experts noted that in handling situations involving violence, the need for a safety plan has the highest priority in protecting victims' safety. Participants noted that the list format often used in LLM responses may imply a ranked importance of information, even though this is not intentional or by design. Even if prioritization were intentional, accurately determining the order of needs often requires years of experience working with the community.

\end{itemize}

\end{document}